# Which Top Energy-Intensive Manufacturing Countries Can Compete in a Renewable Energy Future?

Arne Burdack[1,2,4,*], Maximilian Stargardt[1,2], Christoph Winkler[1,2], Konrad Klein[1], Detlef Stolten[2], Jochen Linßen[1], Heidi Heinrichs[1,3]

[1] Institute of Climate and Energy Systems – Jülich Systems Analysis (ICE-2), Forschungszentrum Jülich, 52425 Jülich, Germany
[2] RWTH Aachen University, Chair for Fuel Cells, Faculty of Mechanical Engineering, Aachen, 52062, Germany
[3] University of Siegen, Chair for Energy Systems Analysis, Department of Mechanical Engineering, 57076 Siegen, Germany
[4] Lead contact
*Correspondence: a.burdack@fz-juelich.de

## CONTEXT AND SCALE

In a world rapidly shifting to renewable power and greenhouse gas-neutral industrial production, the competitiveness of today's top manufacturing countries is being questioned. The key issue is whether energy-intensive processes can continue to be located in today's industrial centers at competitive energy costs in a renewable-powered future. This study uses detailed energy system modeling to quantify the *Renewable Pull—an* economic incentive for industry relocation toward regions with superior renewable resources. Focusing on the chemical, steel, and cement sectors, we quantify the *Renewable Pull* on the ten *top manufacturing countries* and assess how targeted import strategies could mitigate this phenomenon.

## SUMMARY

In a world increasingly powered by renewables and aiming for greenhouse gas-neutral industrial production, the future competitiveness of today's *top manufacturing countries* is questioned. This study applies detailed energy system modeling to quantify the *Renewable Pull—an* incentive for industry relocation exerted by countries with favorable renewable conditions. Results reveal that the *Renewable Pull* is not a cross-industrial phenomenon but strongly depends on the relationship between energy costs and transport costs. The intensity of the *Renewable Pull* varies, with China, India, and Japan facing a significantly stronger effect than Germany and the United States. Incorporating national capital cost assumptions proves critical, reducing Germany's *Renewable Pull* by a factor of six and positioning it as the second least affected *top manufacturing country* after Saudi Arabia. Using Germany as a case study, the analysis moreover illustrates that targeted import strategies, especially within the EU, can nearly eliminate the *Renewable Pull*, offering policymakers clear options for risk mitigation.

## GRAPHICAL ABSTRACT

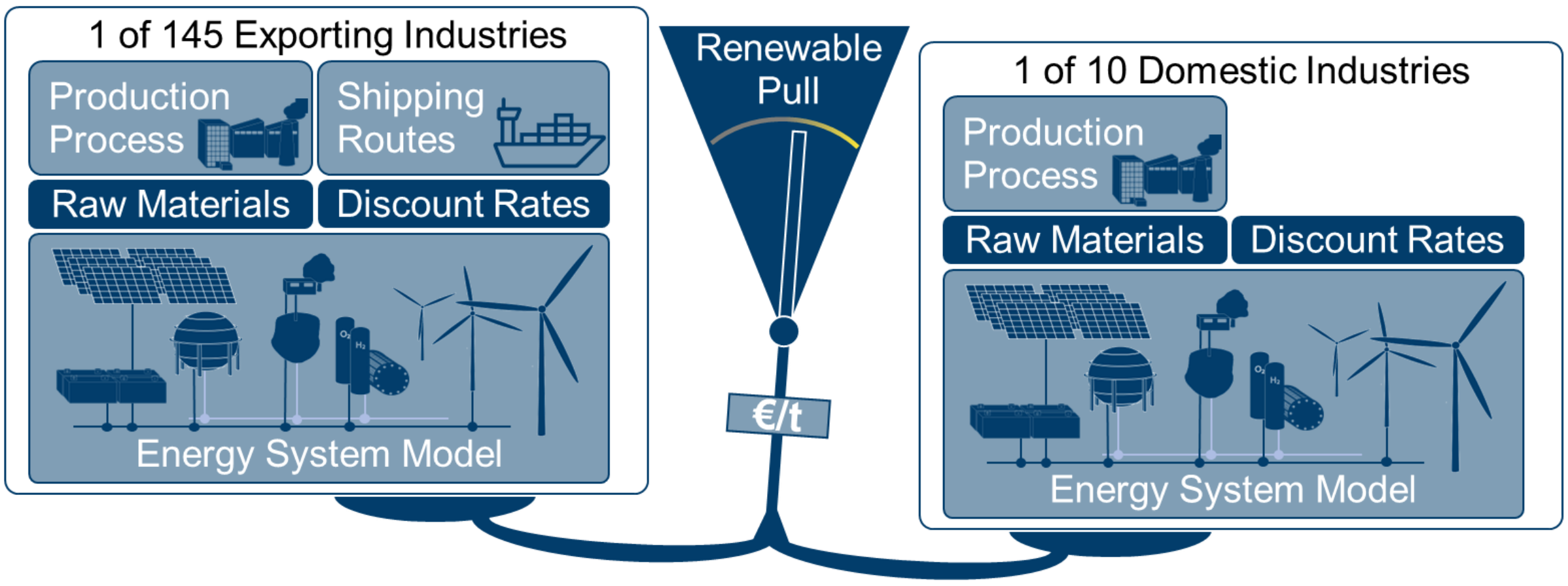

## Highlights

- Nine out of ten top manufacturing countries are exposed to a *Renewable Pull*
- Almost no *Renewable Pull* is observed for the cement industry
- Considering national capital costs, Germany, and the United States face the lowest *Renewable Pull* after Saudi Arabia
- Under uniform capital costs, Saudi Arabia, Iran, and the United States of America face the lowest *Renewable Pull*, while Germany lies in the higher mid-range
- EU imports of hydrogen and intermediates can mitigate the *Renewable Pull* on Germany

## INTRODUCTION

In the context of economic growth and the resulting prosperity for a national economy, the industrial sector stands out from other sectors. As highlighted by the Kaldor-Verdoorn laws—whose underlying mechanisms remain widely recognized—the industrial sector can serve as a nucleus for economic growth [1]. Its importance is further reflected in its share of national value added: In the European Union, the manufacturing sector accounted for 17% of value added in 2018, with Germany leading the way at 22%. In China, the share is as high as 29%, and in the United States, it is 12%. What these economies have in common is that the manufacturing sector is the largest of all economic sectors [2]. The manufacturing sector, particularly its energy-intensive sub-sectors, provides the base materials that form the foundation of long and complex value chains. Beginning with the provision of basic materials, these chains extend through multiple stages of processing and are supported by a wide range of associated services. Intermediate products, employment opportunities, and highly skilled professions are part of it. [3,4] Partly the produced goods are traded on global markets directly. However, base materials such as steel and basic chemicals serve as inputs for the domestic industry, thus representing the starting point for competitiveness across the entire value chain. For example, approximately 25% of global steel production is exported as semi-finished or finished product [5]. Global trade also enables firms to make their location decisions within a globalized context, as evidenced by the relocation announcements of firms compiled by Verpoort et al. and Samadi et al. [6,7]. Among factors such as labor availability, skills and costs, market integration, political framework, and infrastructure, the supply costs of energy and raw materials are important determinants of firms' location decisions [8–11]. Especially for the energy-intensive base industry, the energy costs share in the production make up more than 10% of the production costs and up to 30% of the energy cost share in gross value added [12–14]. It is therefore not surprising that a survey of ten companies from Germany's base industries—representing one of the world's *top manufacturing countries—consistently* identified energy costs as a key location factor [15].

Since the mid-1800s, energy-intensive industries have maintained a supply chain structure based on fossil energy supplies [16]. Due to political action for climate change mitigation and the unprecedented decline in the cost of solar and wind power [17], renewable electricity capacities—especially solar PV and wind capacities—have increased many times over since 2010 [12]. Numerous models have been created to generate scenarios for the future development of electricity generation. Over 3000 model-based climate change mitigation scenarios were compiled as part of the *Sixth Assessment Report* of *The Intergovernmental Panel on Climate Change* (IPCC). Considering all model results based on the *Middle of the Road* scenario [18], solar and wind energy provide an average of 63% of global electricity in 2050, compared to 3% in 2010 [19]. This shift from fossil to variable renewable-based energy systems fundamentally reshapes the spatial patterns of low-cost energy supply, as variable renewables exhibit pronounced spatial cost variations [16,20]. Examining this in the context of today's *top manufacturing countries* defines the research question of this study:

**Can today's *top manufacturing countries* remain competitive in an energy future based on variable renewables?**

A number of recent studies have examined the competitiveness of current, mostly European, industrial locations. In this context, Samadi et al. [7] provide an initial conceptual illustration of the *Renewable Pull* and support it with an exemplary calculation for Morocco and Germany. They use the term *Renewable Pull* to describe an effect "which could lead to a loss of industrial production or a lack of investment in new industrial capacity in certain countries (…) because of superior renewable energy conditions in other countries" [7]. Without explicitly referring to the concept of the *Renewable Pull*, Egerer et al. [21] conduct a comprehensive study based on energy system model calculations, comparing the production costs of steel, ethylene, and urea in seven exemplary countries with excellent renewable potentials to those in Germany. Although domestic electricity demand is not included in the analysis and assumptions on variable capital costs are not country-specific, the study yields notable findings: a substantial cost advantage for relocation, particularly for urea and ethylene, and a clear cost-reducing effect for domestic production from importing intermediates [21]. Nykvist et al. [22] examine the *Renewable Pull* effect on the steel industry using 12 exemplary countries. While they acknowledge the *Renewable Pull* as a relevant concern, they conclude that moderate subsidies, which they frame as a "Renewable Push," could help counteract the *Renewable Pull*. Their analysis is not based on an energy system model but instead relies on levelized cost of electricity (LCOE) data from the literature [22]. Lopez et al. [23] investigate supply chain configurations for cost-optimal provision of e-polyethylene in Europe, based on imports from Morocco and Chile. Electricity costs are estimated using assumed full load hours. The study concludes that domestic production in Europe may come under pressure from low-cost imports [23]. Verpoort et al. [6] provide a valuable conceptual framing of the so-called "green relocation" phenomenon—the potential consequence of a *Renewable Pull*—offering a broader picture by examining the incentives and implications of the *Renewable Pull* from different perspectives. For Germany, they calculate production costs up to 150% higher than those of imports for steel, urea, and ethylene, based on LCOE assumptions for 10 countries and uniform transport cost assumptions. They discuss that offsetting such cost differences through subsidies would require a substantial share of the national budget of Germany. Notably, the chosen range in their capital cost assumptions across regions (5% to 8%) is very small in their analysis [6].

The energy system model plays a central role in quantifying the *Renewable Pull* in this study, as do potential imports in preventing it. Previous studies on imports of goods and energy carriers have high methodological proximity because several of them are based on detailed energy system models used to calculate potential import costs. Therefore, the following studies remain relevant, even if they do not directly address the industrial sector or the *Renewable Pull* itself. Hampp et al. [24] focus on large-scale renewable energy carrier supply for Germany, analyzing eight countries using the energy system modeling framework PyPSA, with detailed representations of wind and solar potentials. In contrast to the previously mentioned studies, domestic energy demand in the modeled regions is accounted for by subtracting it from the available potential prior to the optimization, meaning the domestic electricity demand is not part of the optimization. The study concludes that Germany has several cost-effective import options for renewable energy carriers—an aspect that is highly relevant in the context of this study [24]. Also based on PyPSA but extended with a global energy supply chain model and centered around a European multi-node, sector-coupled energy system model, Neumann et al. [25] published a preprint analyzing the impact of energy carrier and industrial imports on the European energy system costs. Focusing on the European domestic energy system, they find that imports can reduce total system costs by 1% to 14%. The resulting discussion, whether such comparatively modest cost advantages should be foregone in favor of domestic value creation and job generation, closely aligns with the motivation behind investigating the *Renewable Pull* for energy-intensive industries within this study [25].

This study aims to connect detailed energy system modeling with the existing literature on industrial competitiveness and the *Renewable Pull* and to contextualize the phenomenon on a global scale. This global perspective is essential for moving beyond national case studies. It provides an overarching view of its global relevance while enabling a differentiated assessment of how strong individual countries are affected. We apply detailed energy system modeling, shipping, and transmission calculations to quantify the *Renewable Pull* and the effectiveness of countermeasures in the future. The aim is to obtain highly differentiated and robust results on a global scale, which is made possible through the use of most recent and well-acknowledged renewable potential data. By analyzing the *Renewable Pull* on the world's 10 *top manufacturing countries—considering* every country of the world having a port as potential export country—this study expands the current state of research by examining the *Renewable Pull* in a global context.

The analysis begins with the energy system model-based calculation of production costs per country. In contrast to previous studies, where the industrial sector is modeled separately from the domestic energy system, this study integrates the industrial sector directly into each country's energy system, which by that must also meet the electricity demands of other sectors. The optimization of the production processes and the domestic power system is treated as a single optimization problem. This ensures that neither industry nor the domestic power sector receives preferential access to the best renewable resources, resulting in the emergence of an overall energy system perspective for each country. Thereby, national assumptions for discount rates ensure a highly differentiated consideration of capital costs, while additionally calculating each scenario with constant discount rates shows the purely energy system-related differences in the production costs. By combining the optimized production costs with a detailed shipping route tool, import costs are calculated for each of the 1 450 export-import country combinations, which allows the calculation of 145 *Renewable Pull* values for each import country. In the final step, the potential of importing the intermediates hydrogen, methanol, and hot briquetted iron is examined as a countermeasure to the *Renewable Pull*, using Germany as an exemplary case. In this context, an EU pipeline and truck import model is applied in addition to the global shipping model, enabling the quantification of how the EU as a single market can mitigate the *Renewable Pull*.

## RESULTS

The core of this study is the quantification of the *Renewable Pull* for each country on a global scale. Therefore, the domestic production costs of olefins, steel, ammonia, and cement have been calculated for the 10 *top manufacturing countries—defined* in the methodology as the top-producing nations of the industrial goods analyzed—based on optimized renewable energy systems. These costs are then compared to the costs of importing the same goods from 145 foreign countries, for which the production costs are calculated in the same way.

### *Energy system modeling is essential: no universal formula for lowest-cost energy systems*

*Many roads lead to Rome,* and many energy system compositions can lead to competitive production costs: overall, no single renewable energy system cost composition emerges as fundamentally superior. In the five lowest-cost cases shown in Figure 1, all considered potentials—onshore wind, solar PV, and offshore wind—are present, and both countries with and without salt cavern storage potential appear. This demonstrates that renewable energy costs cannot be reduced to simplifying assumptions such as the full load hours of individual technologies; instead, a comprehensive energy system framework—as applied in this study—is essential to capture the complex interdependencies.

To provide a fundamental understanding of the optimized energy systems, the first step presents and compares the cost compositions of the energy systems in the 10 *top manufacturing countries* and in five selected additional countries with particularly favorable renewable potentials. Differences in cost composition are analyzed and explained. Figure 1 illustrates the cost components of optimized energy systems for the production of one megaton of olefins per year. Due to their high energy intensity, olefins are the most cost-intensive product considered in this study. The corresponding energy systems are simultaneously required to meet domestic electricity demand. In most countries, more than 90% of the total production costs stem from the renewable power and hydrogen system. Overall, production costs vary by up to 50% across most countries. An exception is South Korea, where costs are more than three times higher than in the lowest-cost countries. Limited onshore wind and solar potential—due to low land eligibility and low achievable full load hours—force a high share of more investment-intensive offshore wind. Yet the main advantage of offshore wind, high full load hours, is absent in South Korea, with a theoretical average of only 2549 compared with 3495 in Germany as an exemplary reference. Moreover, South Korea lacks salt cavern storage and is therefore dependent on more costly hydrogen gas vessels as storage option. The same applies to Japan.

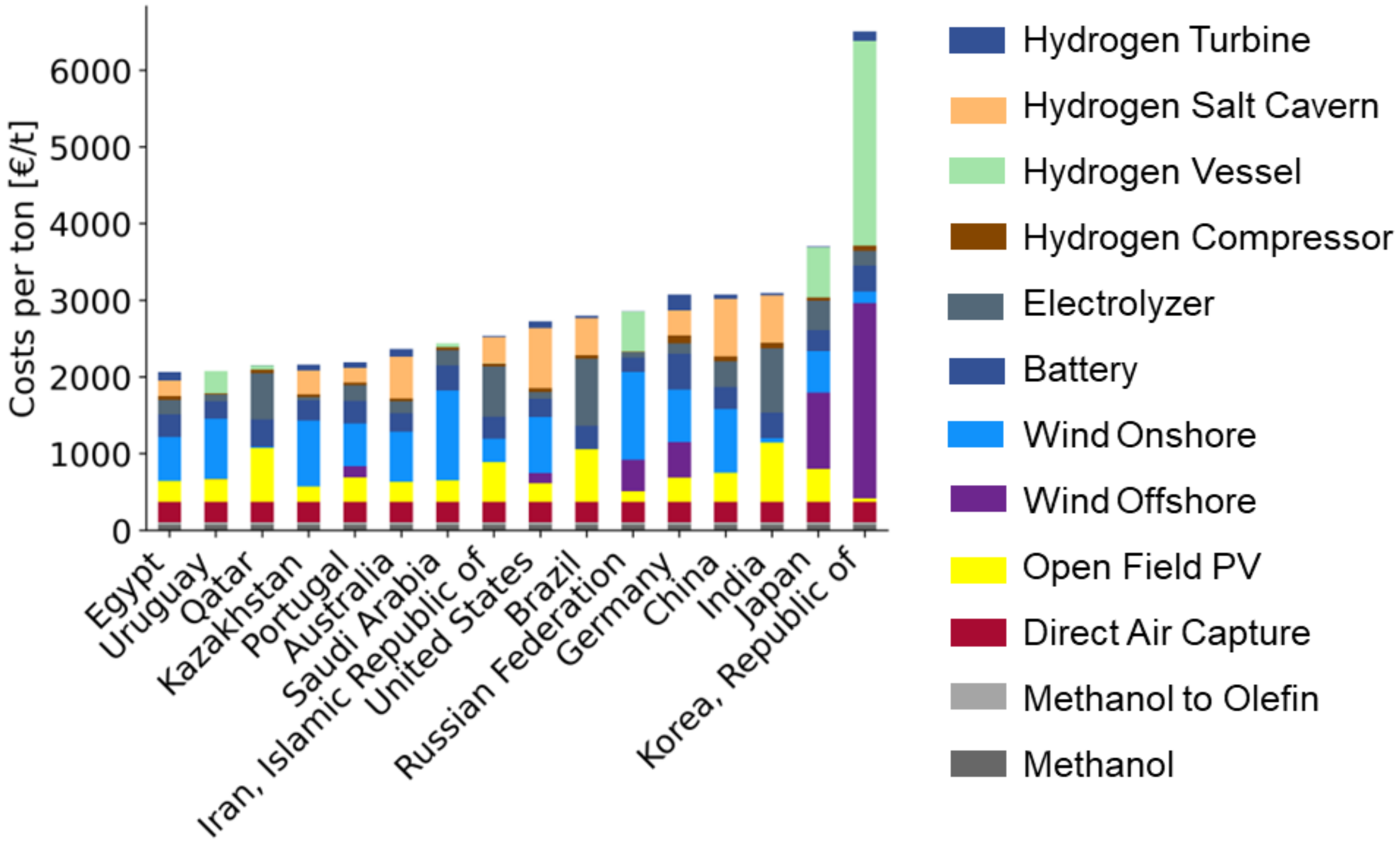

***Figure 1: Cost composition of optimized renewable energy systems for the olefin production of 1 Mt/a in the top 10 manufacturing countries and five low-cost countries from different global regions, considering uniform capital costs***

Japan and South Korea share the absence of salt cavern storage potential with Uruguay, Qatar, and Russia. However, Uruguay and Qatar are able to compensate—Uruguay through low domestic electricity demand and both through excellent renewable potentials. Countries with lower production costs typically feature energy mixes with little to no offshore wind. In contrast, Germany, Portugal, the United States, and Russia include offshore wind in their energy systems due to high full load hours—averaging 3495, 2742, 3366, and 3122, respectively, across all clusters—or, in the case of Portugal, because system-supporting time series help balance onshore wind and solar fluctuations. In Germany's case, certainly also limited overall onshore and solar potentials play a role. With the exception of Brazil and Qatar, most countries utilize hybrid systems. Brazil's purely solar-based system—explainable by low potential onshore wind full load hours averaging only 1118—requires substantial electrolyzer and hydrogen storage capacity, unlike Qatar, which relies more on battery storage and minimal hydrogen storage. Interestingly, in some countries, such as Saudi Arabia, renewable energy capacity is expanded to avoid the need for costly storage solutions beyond batteries. In other countries, such as Brazil, curtailment is minimized using cost-efficient hydrogen salt cavern storage and incurring high electrolyzer capacities, which enable the system to balance the volatile solar PV input.

### *Nine out of ten top manufacturing countries are exposed to a Renewable Pull*

The higher the *Renewable Pull—as* defined in the methodology—the greater the cost advantages of imports over domestic production. As Figure 2 reveals, the *Renewable Pull* is not a phenomenon affecting only a few individual nations but rather a global one. Wind speed and solar irradiation determine the intensity of the *Renewable Pull*, as does the availability of land for renewables, tending to make densely populated, smaller countries more affected than larger ones. This results in Germany being exposed to the third-highest *Renewable Pull* from energy system costs alone, after Japan and South Korea. At the same time, India, despite its size, experiences a *Renewable Pull* comparable to Germany due to its poor wind conditions. In addition to cost differences arising from renewable energy systems, capital costs play a decisive role in the calculated production costs and can even surpass the influence of the renewable energy system costs. As a result of considering national capital costs, Germany—despite its limited land area and high population density—becomes the country with the second-lowest *Renewable Pull*, while Iran—despite its excellent renewable potential—is exposed to one of the strongest.

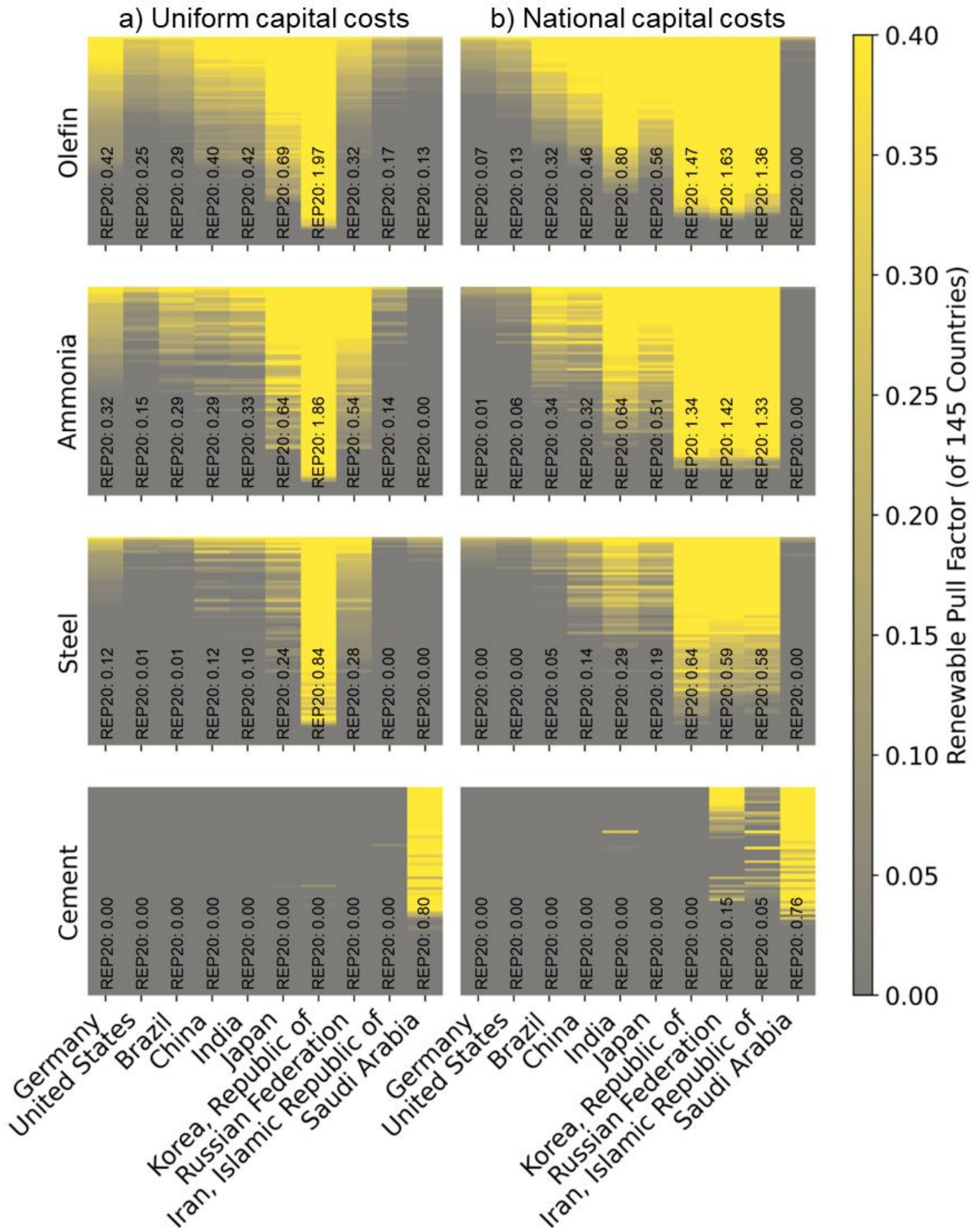

***Figure 2: Renewable Pull among today's leading 10 energy-intensive industry producers Regionalization of the Renewable Pull for the olefin production***

A *Renewable Pull* above 0 indicates that imports are cheaper than domestic production, while a *Renewable Pull* value of 1 corresponds to an import of the respective country costing only half as much as domestic production. To derive all possible *Renewable Pull* values for each top manufacturing country, the exemplary energy system models are combined with shipping costs for every country combination, resulting in 145 *Renewable Pull* values per top manufacturing country. Figure 2 provides a consolidated overview of the intensity and quantity of the *Renewable Pull* on the current *top manufacturing countries*. A distinction is made between calculations with uniform capital costs, reflecting the pure impact of the energy system, and calculations with national capital cost assumptions, which visibly have a significant influence on the results. Assuming uniform capital costs across all countries, the analysis reveals that each of the ten *top manufacturing countries* is exposed to a *Renewable Pull* for at least one of the modeled industrial products. The magnitude of the *Renewable Pull* is closely linked to the energy intensity of the respective processes. The olefin production, being the most energy-intensive process analyzed, incurs the highest production costs. These high production costs not only intensify the effect of regional differences in renewable energy system costs but also reduce the relative impact of transport costs. For olefin production, shipping accounts for less than 4% of the total import costs. In contrast, cement production shows the opposite pattern: its lower process energy intensity results in lower production costs, leading to shipping costs that can make up to 50% of the total import costs, depending on the export country. Consequently, virtually no *Renewable Pull* is observed for cement, with the sole exception of Saudi Arabia, where high limestone costs of €125 per ton drive the effect (see methods for cost approach). In contrast, olefins exhibit a pronounced Renewable Pull, with an average REP_20 of 0.68 across all country combinations. REP_20 refers to the Renewable Pull threshold that is exceeded by 20 countries. The *Renewable Pull* is particularly high in countries with limited land availability for renewables, such as South Korea, Japan, and Germany—the three smallest countries by area among the *top manufacturing countries*. Among them, South Korea exhibits the highest *Renewable Pull*, driven by its high population density and mountainous terrain, which constrains the usable renewable potentials [26]. Furthermore, both South Korea and Japan lack salt cavern storage capacity and must rely on expensive hydrogen gas vessel storage, adding substantially to the overall system costs—reaching the level of the total calculated production costs for the United States.

Introducing national capital cost assumptions into the modeling framework significantly changes the magnitude of the *Renewable Pull*. While it disappears entirely in some countries due to favorable capital costs, it increases severalfold in others. In the olefin model, for example, applying national capital costs reduces Germany's *REP_20* indicator by a factor of six, eliminates the *Renewable Pull* entirely for Saudi Arabia, and increases it eightfold for Iran. Under these assumptions, Germany ranks as the second least affected country and only Saudi Arabia demonstrates a lower *Renewable Pull*. In contrast Russia and Iran continue to show high *Renewable Pull* values, indicating that their competitiveness is challenged even when favorable renewable potentials exist. China and India show *Renewable Pull* values comparable to Germany when using uniform capital costs. However, their *REP_20* values increase up to fourfold when national capital costs are introduced. In addition to higher risk premiums embedded in capital markets, this shift is also due to increasing exposure to natural hazard risks in these regions, an aspect explicitly accounted for in the national capital cost methodology (detailed description in the methodology section). The comparison of results under both capital cost scenarios highlights which countries are most sensitive to shifts in investment conditions. Rising capital costs particularly affect the competitiveness of Japan and Germany. Conversely, countries such as Russia and Iran, despite favorable resource potentials, face a significant *Renewable Pull* due to high national capital costs.

### *30 countries exerting a Renewable Pull on Germany*

Following the illustrative overview in Figure 2 on the intensity and extent of the *Renewable Pull* affecting the *top manufacturing countries*, the next question is where this pull originates for each of them. The patterns differ for every *top manufacturing country* and are influenced by the shipping costs for transporting goods from various regions. As described, the intensity of the *Renewable Pull* correlates with the energy intensity of the industrial process. Across the different processes examined, similar regional patterns of the *Renewable Pull* emerge for each top manufacturing country. Therefore, the geographically resolved plots are shown for the olefin industry, as this is where the highest *Renewable Pull* values occur, for the three largest producers in the study—China, the United States, and India—alongside Germany, which is the representative of the EU among the *top manufacturing countries*.

With uniform capital costs, Germany experiences a *Renewable Pull* from a diverse range of regions across all continents, including the Iberian Peninsula. The strongest pulls, reaching 0.7 and above, originate from the African continent, while the *Renewable Pull* from the Arabian continent, Uruguay, and Kazakhstan is around 0.4. Moderate *Renewable Pull* values of approximately 0.24 come from

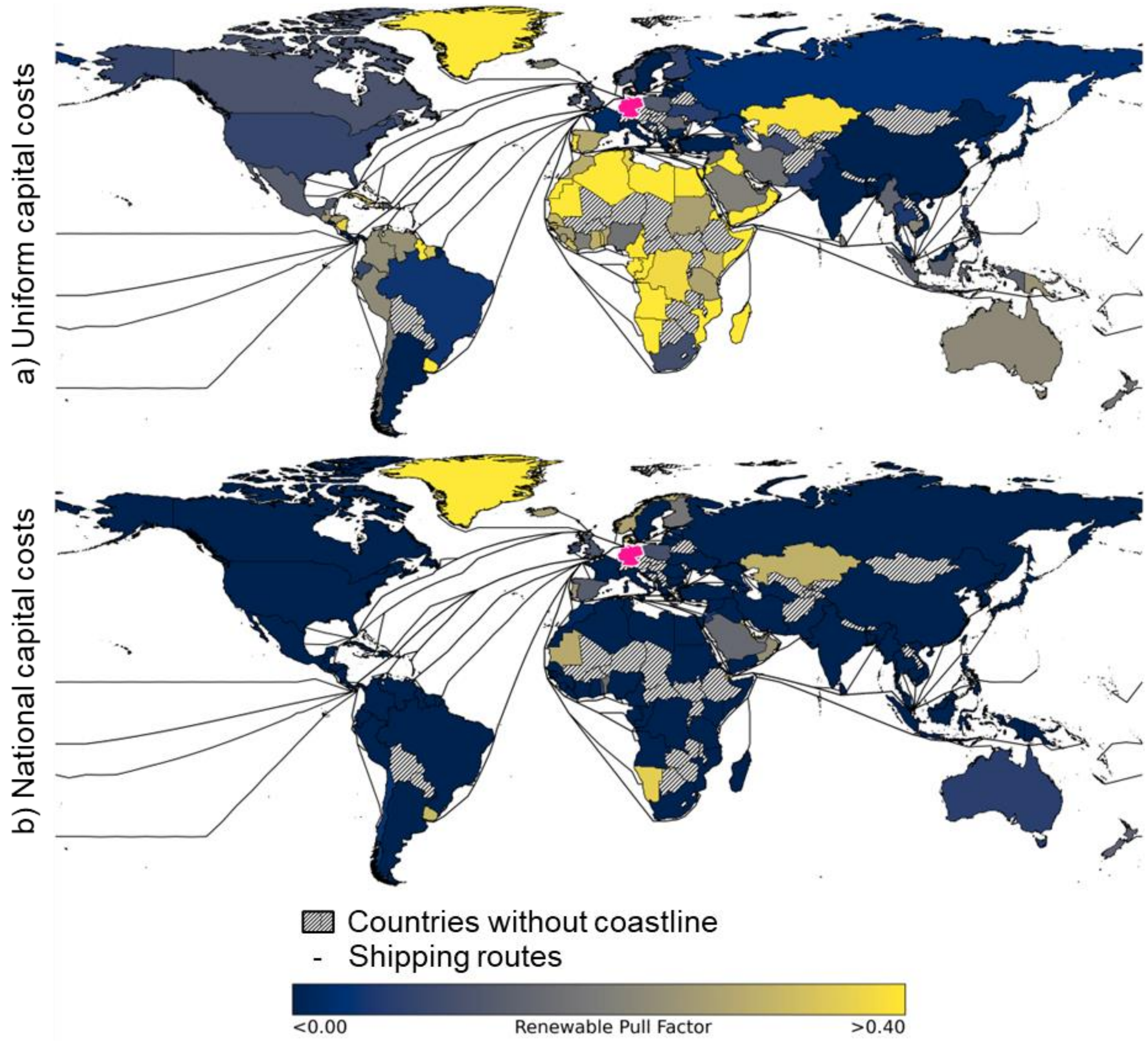

***Figure 3: Renewable Pull origin and intensity for the olefin production of Germany***

Australia and Latin America, and about 0.1 from the United States. No *Renewable Pull* is observed from China or India.

When national capital costs are applied, the *Renewable Pull* on Germany comes from 30 regionally distributed countries. The strongest values are from Qatar (0.36), Kuwait (0.26), and Namibia (0.24). Denmark, Uruguay, Kazakhstan, Mauritania, and Norway each exert a pull of around 0.2, while all other regions with a *Renewable Pull* remain below 0.16. Notably, no *Renewable Pull* originates from East Asia or the United States.

***41 countries exerting a Renewable Pull on the United States***

When uniform capital costs are considered, the United States experiences the strongest *Renewable Pull* from parts of Africa, with values reaching up to 0.62 from Western Sahara, 0.51 from Mauritania, and 0.42 from Namibia. *Renewable Pull* values between 0.2 and 0.3 originate from the Middle East, Kazakhstan, Uruguay, Iraq, and Portugal. No Renewable Pull is observed from China or India.

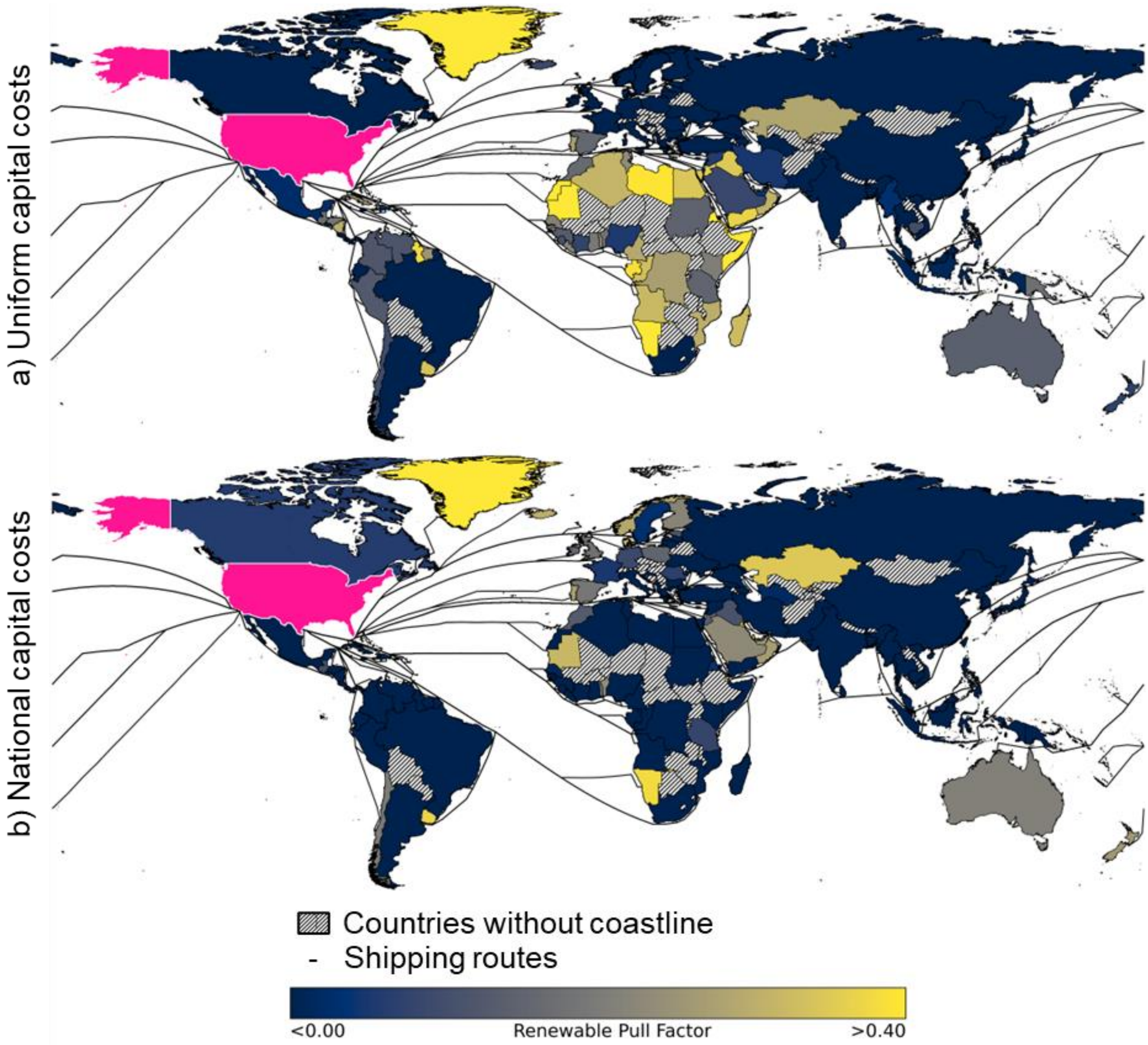

***Figure 4: Renewable Pull origin and intensity for the olefin production of the United States of America***

When national capital costs are applied, the *Renewable Pull* on the United States comes from 41 countries. The highest values arise from Qatar (0.42), Kuwait (0.32), Namibia (0.31), and Uruguay (0.29). Several EU countries also exert a notable *Renewable Pull*—Denmark (0.26), Norway (0.21), and Portugal (0.21)—each exceeding the *Renewable Pull* from Australia (0.13). No *Renewable Pull* is observed from East Asia, and beyond Uruguay, only low values occur from Latin America.

### *72 countries exerting a Renewable Pull on China*

When uniform capital costs are considered, China experiences the strongest *Renewable Pull* from the African continent, with values of 0.78 from Western Sahara, 0.72 from Eritrea, and 0.60 from Namibia, followed by the Middle East with 0.48 from Yemen and 0.44 from Oman. Interestingly, the *Renewable Pull* from Portugal (0.34) is stronger than from Australia (0.28). Kazakhstan (0.36) and Uruguay (0.41) also show high values, although Kazakhstan's *Renewable Pull* is likely underestimated since only shipping was considered in its calculation, while the two countries share a common border. The United States exerts a comparably moderate *Renewable Pull* of 0.10.

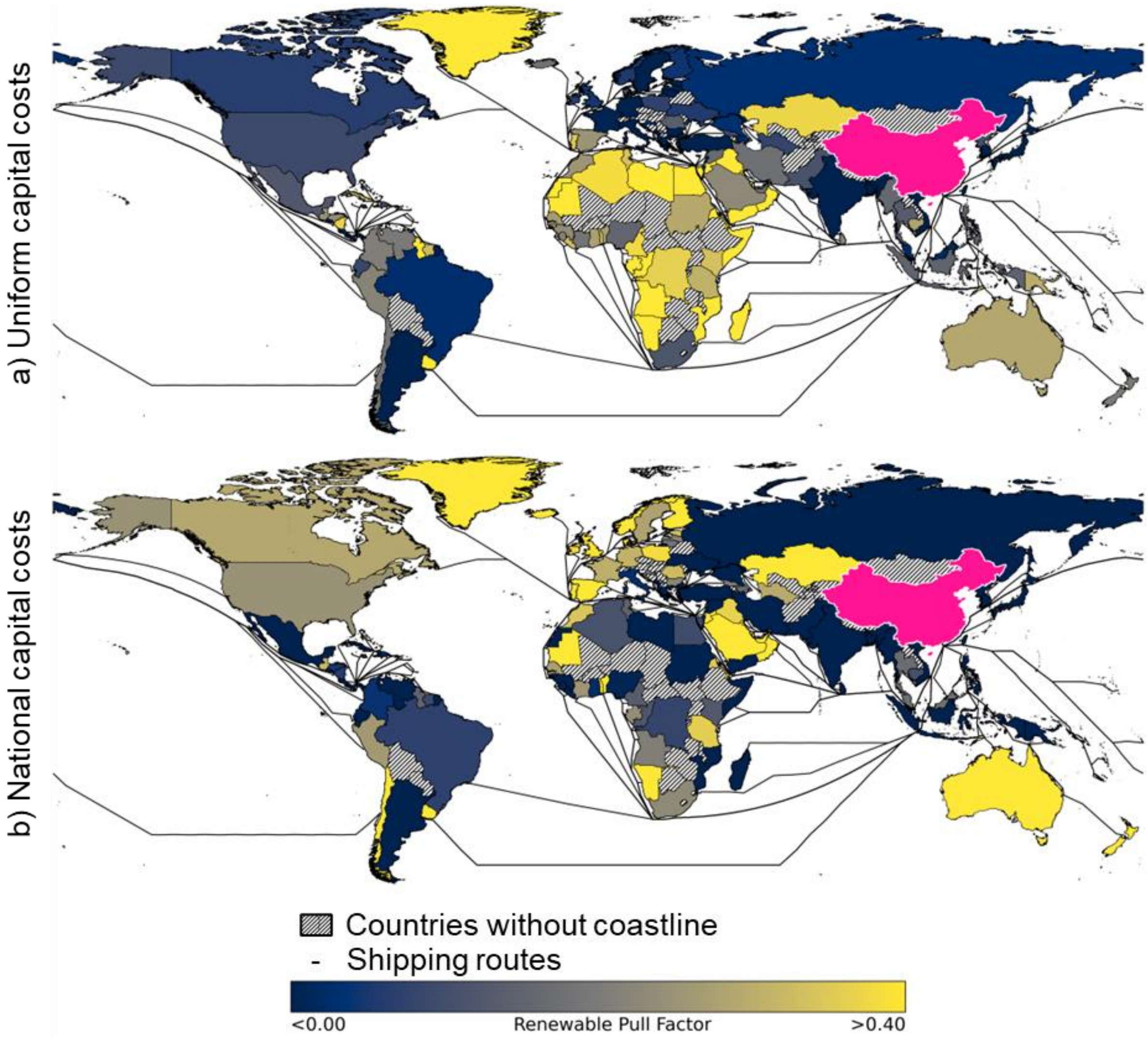

***Figure 5: Renewable Pull origin and intensity for the olefin production of China***

When national capital costs are applied, the number of countries exerting a *Renewable Pull* on China increases to 72, which is an order of magnitude higher than for the United States or Germany. The intensity also rises significantly, reaching 0.88 from Qatar, 0.75 from Kuwait, and 0.68 from Namibia. Multiple European, African, and Middle Eastern countries exceed a *Renewable Pull* of 0.5 on China, while Australia comes close at 0.48, which is considerable, given that a *Renewable Pull* of 0.5 means domestic production costs are 1.5 times higher than import costs. Other *top manufacturing countries* also exert a notable *Renewable Pull* on China, including the United States (0.25), Germany (0.30), and Saudi Arabia (0.52).

***97 countries exerting a Renewable Pull on India***

With uniform capital costs, the picture for India resembles that for China. However, when national capital costs are applied, the *Renewable Pull* becomes even more intense across all continents, including the EU and the United States, and exceeds the values observed for China. Already burdened by comparatively unfavorable capital costs from a financial market perspective, India's rates are further elevated by the projected likelihood of future natural hazards. In total, 97 countries exert a *Renewable Pull* on India. The highest values are found for Qatar (1.35), Kuwait (1.18), and Namibia (1.07), where domestic production costs exceed import costs by more than a factor of two. With the exception of Russia and Iran, all other *top manufacturing countries* exert a *Renewable Pull* on India.

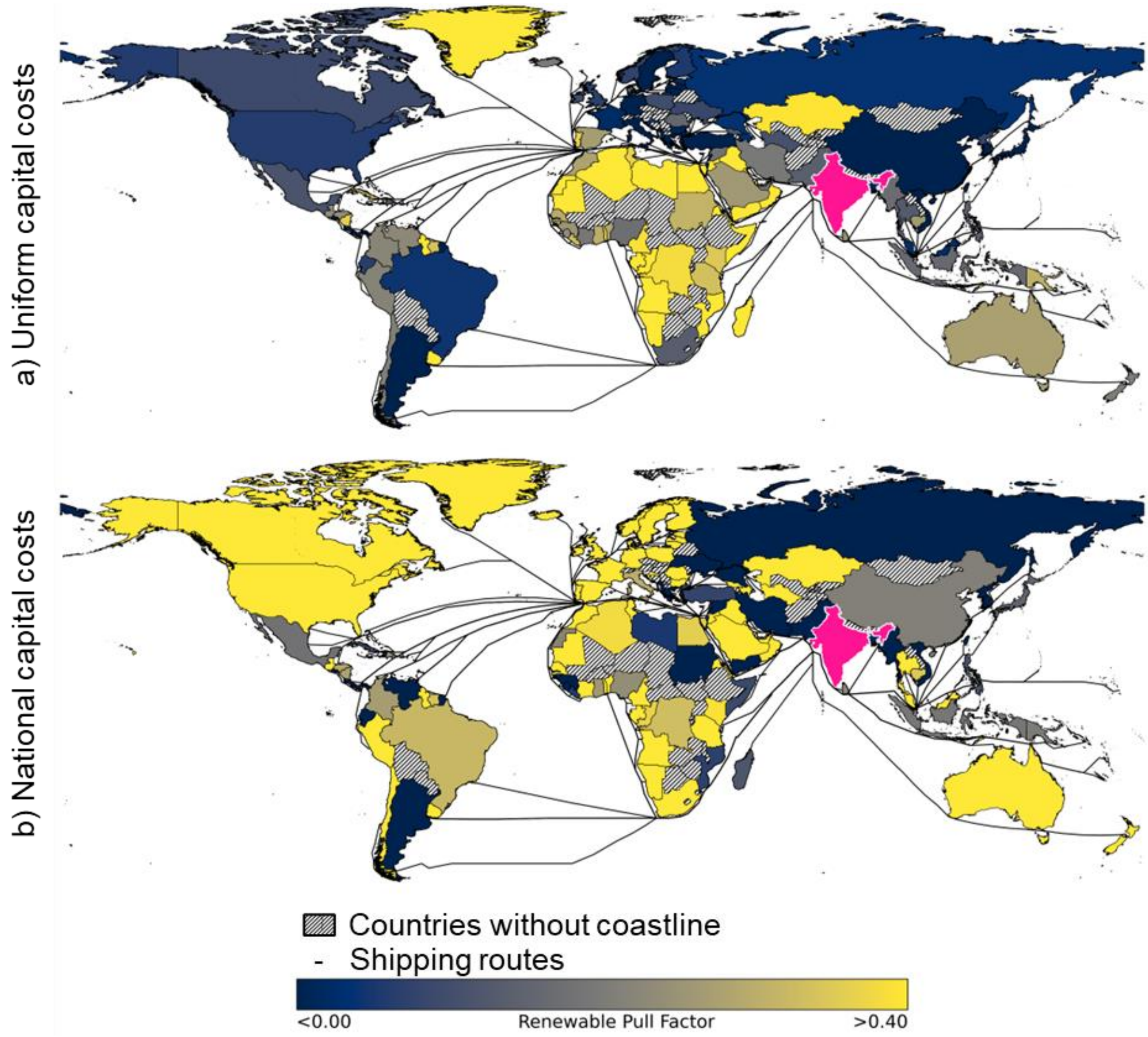

***Figure 6: Renewable Pull origin and intensity for the olefin production of India***

With maximum specific transport costs of 120 EUR/t, shipping routes have a limited impact on the *Renewable Pull* for Olefin production, where production costs for most countries range between 2 000 EUR/t and 3 000 EUR/t. In this setting, transport costs can at most offset a *Renewable Pull* of 0.05 that would occur without considering transportation. By contrast, in the steel industry, transport costs can offset a *Renewable Pull* of up to 0.13, and in the cement industry even up to 1. The lower the specific total production costs, the greater the protection provided by transportation costs against an energy-driven *Renewable Pull*.

*EU hydrogen and intermediate imports can largely mitigate the Renewable Pull on Germany*
Having established that all top manufacturing countries, except Saudi Arabia, will face a *Renewable Pull* in a renewable-powered world, the final part of the results investigates countermeasures to be able to keep industries competitive, even if a *Renewable Pull* occurs. Therefore, the potential to counteract the *Renewable Pull* effect through targeted import strategies is examined. Germany is selected as an exemplary case for two main reasons: first, as a geographically small country, it will inevitably have to rely on import strategies in the future [27]. Second, as a member of the EU, it benefits from being part of a large single market, where fewer barriers for multinational infrastructure development and on truck routes are expected compared to countries that are not economically integrated in a similar way. Based on an EU import optimization model, whose results are shown in Figure 7 (Model 2, Methods, Figure 10 for details), and a global ship-based import model (Model 1, Methods, for details), global ship import costs and EU pipeline and truck import costs have been calculated for Germany.

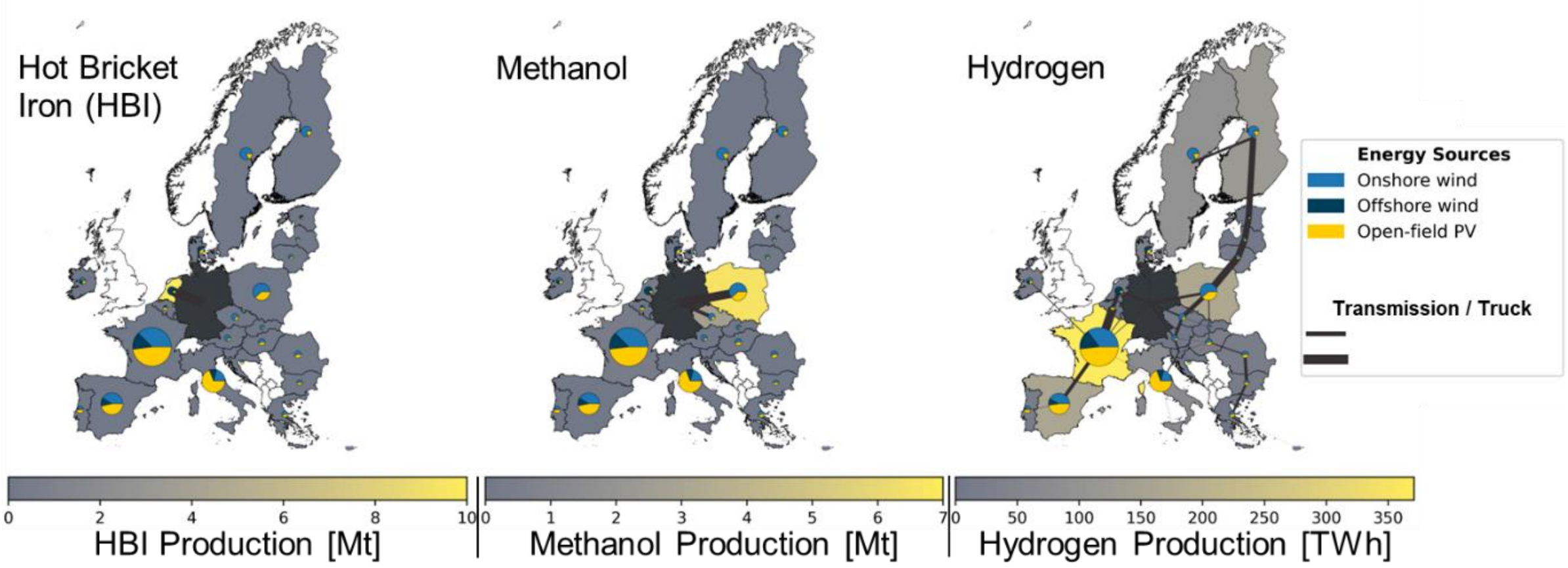

*Figure 7: Optimized EU import infrastructure for German imports under national capital costs and considering 100% of domestic electricity demand*

Building on the calculated import costs, the German optimization model could then choose between EU pipeline and truck imports or global shipping imports. Figure 8 presents the results in three stages: production costs in Germany without imports, production costs with imports from around the world (including EU pipeline and truck imports), and production costs when the model is restricted to EU imports only. For steel production, an additional stage is included, allowing for the import of HBI. The resulting production costs are compared with those of countries among the range of the 20 exerting the strongest *Renewable Pull* on Germany (REP_20).

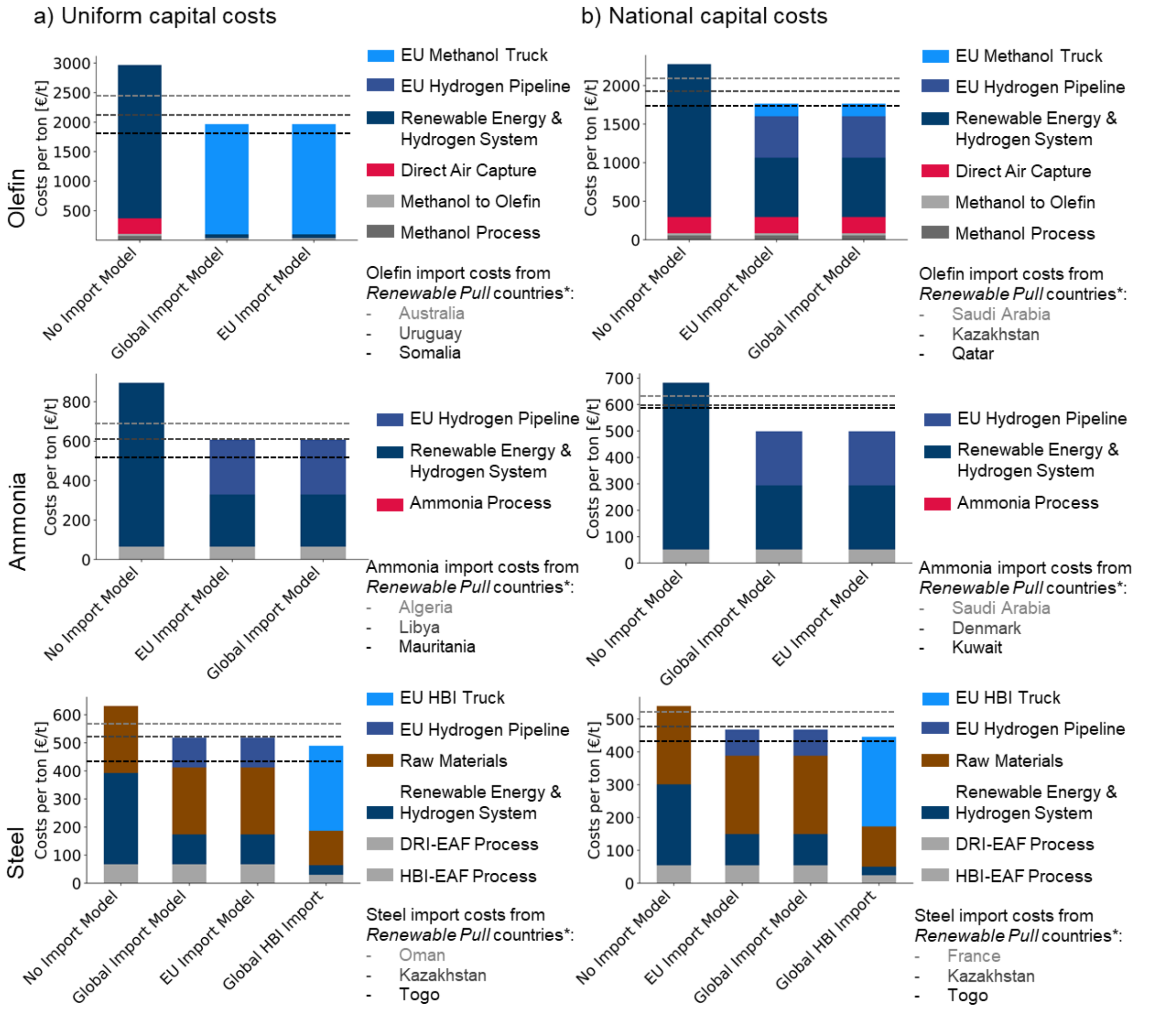

***Figure 8: Optimized production cost reduction potentials through imports for Germany under different import options, left side: a) based on uniform capital costs, right side: b) based on national capital costs, *Countries mapping the range of the 20 strongest Renewable Pull countries (REP_20)***

For both olefin and ammonia production, Germany can largely avoid the *Renewable Pull* by importing methanol and hydrogen. Under uniform capital costs, the cost-optimal solution results in no domestic methanol production for olefin manufacturing. In contrast, when national capital costs are applied, all methanol process capacities (100%) for olefin production are built domestically. In steel production, Germany can also largely avoid the *Renewable Pull* by importing either hot briquette iron or hydrogen. While HBI imports almost eliminate the *Renewable Pull*, they do so at the expense of domestic iron production capacities. Alternatively, importing hydrogen significantly reduces the *Renewable Pull* while preserving domestic iron production from iron ore. Across all import scenarios, the model consistently favored EU imports over global shipping due to lower costs. In short, Germany can largely avoid the *Renewable Pull*, with the most cost-effective imports coming from EU partners via truck or pipeline.

## DISCUSSION

National capital cost assumptions can fundamentally reshape the resulting *Renewable Pull*. Within this study, four factors shape the intensity and quantity of the *Renewable Pull* on each country: renewable energy potentials, salt cavern storage potentials, capital costs, and shipping routes. Figure 1 shows differences in the energy system cost compositions. Figure 12 shows exemplary the different transport costs. Beyond that, national capital costs stand out as especially influential. In this analysis, the assumed national capital costs are derived 75% from current financial market data and 25% from projected increases in natural hazard risk, following the most recent approach of Stargardt et al. [28]. India, with both high hazard exposure and comparatively high market-based capital costs, shows a significant jump in the *Renewable Pull* when national capital costs are applied. China, Japan, and the United States are also affected. Germany presents the opposite case: already low capital costs fall slightly further due to minimal hazard risk. Capital costs merit this differentiated treatment, yet they are not physically grounded like the energy system modeling. Instead, they rest on estimates of volatile market dynamics and future natural hazard probabilities. Government subsidies or sudden changes in investment capital costs could therefore alter or lower the *Renewable Pull*. Still, the key added value of this study remains: identifying the physically grounded cost differences in the renewable energy system and contextualizing the influence of national capital costs relative to renewable energy system–driven production cost differences on a global scale. Among the *top manufacturing countries,* converging capital costs would, as Figure 1 shows, benefit Iran, India, Russia, and Brazil most, while the *Renewable Pull* from the 20 countries exerting the strongest effect on Germany (REP_20) would increase sixfold.

The results prove that the *Renewable Pull* primarily affects highly energy-intensive industrial sectors with low transport costs compared to the overall production costs, while the cement sector remains unaffected. Consequently, potential measures to mitigate the *Renewable Pull* should be tailored to the particularly affected sectors rather than be spread across the entire industry.

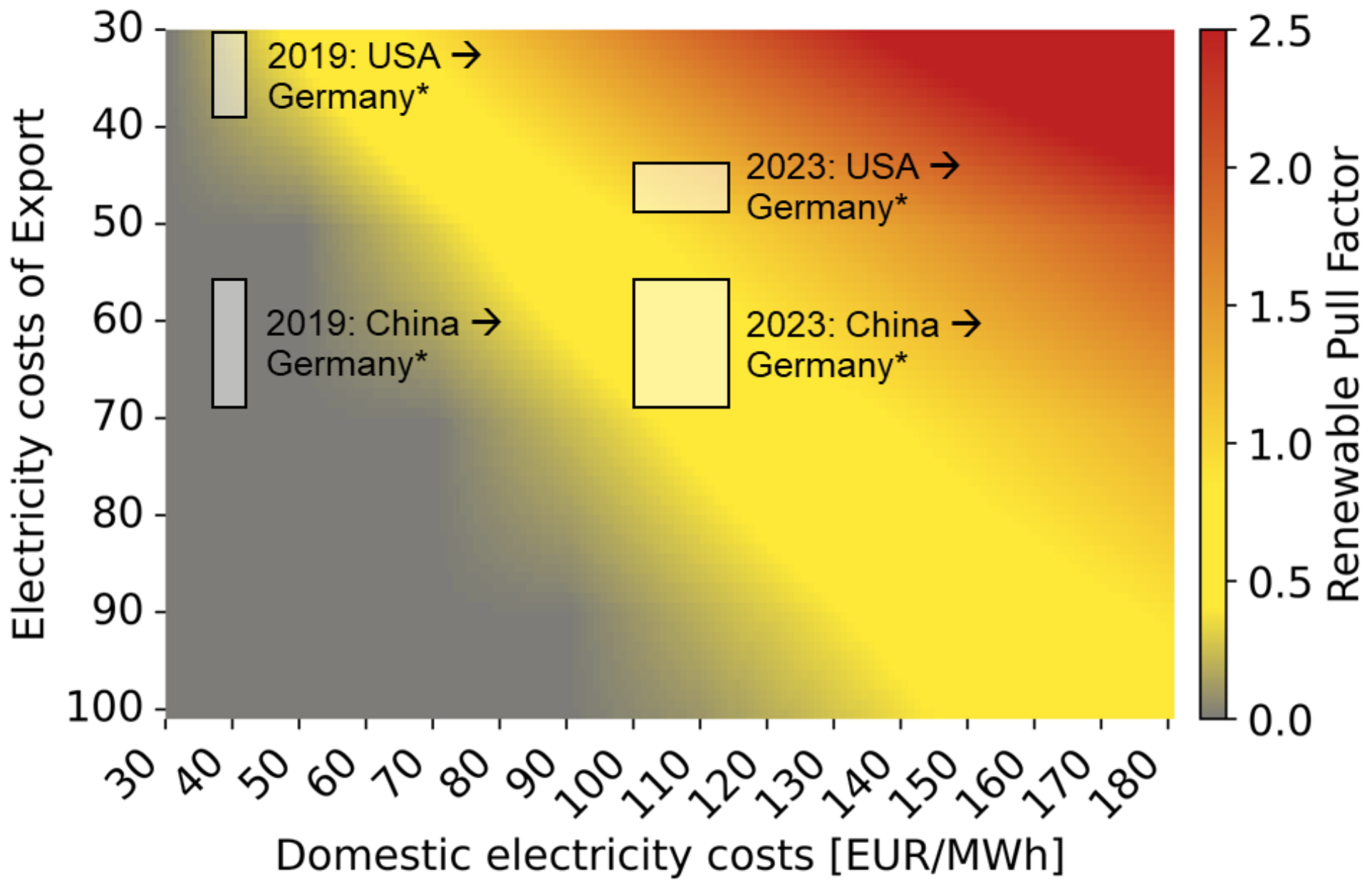

***Figure 9: Renewable Pull Factor for the Olefin Industry as a function of electricity cost differences between two countries, *the boxes indicate the range of theoretically calculated Renewable Pull between two countries, based on electricity prices for energy-intensive industries in 2019 and 2023 [29], → the arrows indicate the direction of the Renewable Pull***

If the *Renewable Pull* were calculated based on current electricity prices, it would reach higher values for European countries than the maximum values derived from renewable energy system cost differences. Figure 9 illustrates a simplified *Renewable Pull* matrix calculated solely from electricity cost differences. It is therefore not directly comparable to the complex energy-system-model-based calculations presented in this study but rather serves as an approximate way to assess the pull effects arising from current electricity prices. Gray and yellow shades indicate increasing *Renewable Pull* factors with rising price gaps; once red tones appear, the maximum *Renewable Pull* values from this study are exceeded. The diagonally aligned yellow band from the upper left to the lower right represents the range of *Renewable Pull* values between 0 and 0.9 expected for 2050 for the olefin process based on this study (excluding the extreme cases of Japan and South Korea, which face maximum values between 2 and 3). A simplified experiment illustrates the implications of current electricity price differences when placed in the *Renewable Pull* matrix shown in Figure 9. In 2019, electricity prices for Germany's energy-intensive industry ranged from 38–42 EUR/MWh, compared to 30–39 EUR/MWh in the United States and 58–69 EUR/MWh in China [29]. By 2023, German prices had risen to 101–114 EUR/MWh, compared to 44–49 EUR/MWh in the United States and 55–69 EUR/MWh in China [29]. While this implies little to no pull on Germany in 2019, for 2023 the Germany–United States electricity price gap yields theoretical pull values exceeding 0.9 for Germany and between 0.1 and 0.9 for the Germany–China electricity price gap. This demonstrates that pull effects based on fossil energy supply can exceed those calculated for renewable energy systems. It follows that, while this study addresses the future competitiveness of industry, for some top manufacturing countries it must first be ensured that their energy-intensive sectors can withstand today's fossil-based pull effects.

This study discusses the *Renewable Pull* as a phenomenon that, to some extent, presupposes a world order based on free trade. Government tariffs can counteract the effect of the *Renewable Pull* just as subsidies can support domestic production. In recent years, however, the announcement of various—sometimes drastic—tariffs by the United States against different countries has added another variable to the discussion of location competitiveness [30]. A detailed discussion of the impacts of such tariffs lies beyond the scope of this study, but such tariffs can be set in relation to an according *Renewable Pull* value: a 15% import tariff offsets a *Renewable Pull* of approximately 0.18, a 30% tariff corresponds to 0.43, and a 50% tariff to 1.

It has been revealed that imports can largely mitigate the *Renewable Pull*. The import modeling results for Germany show that a targeted strategy for hydrogen, methanol, and hot briquette iron can largely prevent the *Renewable Pull*, even under uniform capital costs. While hydrogen imports would, in terms of lost domestic value creation, resemble today's fossil fuel imports, methanol and hot briquette iron imports would reduce the domestic value added. This creates a trade-off between such losses and the risk of downstream industry relocation. To the authors' knowledge, a precise quantification of value-added losses per process step is not available within public data. However, in steel production we know that value creation from processing exceeds that from primary production [2]. If high-quality steel could still be manufactured domestically from imported hot briquette iron, the most lucrative part of the value chain would remain. A similar pattern is conceivable for methanol imports in the chemical sector.

Another advantage of imports is the possibility of relieving the domestic energy system. A sensitivity analysis for the olefin industry (Figure S4 15) shows that a 200 TWh increase in electricity demand can raise production costs by 18%, while an equivalent decrease can lower them by 6%. With German olefin production of about 5 Mt/year [31], the methanol-to-olefin route alone would require 150 TWh for methanol production if 100% of olefins were produced via this pathway—a simplifying assumption. Thus, a combination of energy system relief and retention of most value creation appears feasible.

The results of this study provide a comprehensive global assessment of the *Renewable Pull* introduced in previous literature. It is now evident that the *Renewable Pull* is indeed a global phenomenon, affecting 9 out of 10 top manufacturing countries. Extending the predominantly Europe-focused perspective of earlier studies, or the focus on single exemplary countries of different world regions—as Nyquist et al. [22] did—to the top 10 manufacturing countries in the world has therefore proven valuable. Germany, as an exemplary region of Europe, is found to be subject to a comparatively low *Renewable Pull*. Scenarios by Verpoort et al. [6], in which Germany would need to invest a substantial share of its national budget in subsidies to counteract the *Renewable Pull*, could not be confirmed. Instead, targeted imports of hydrogen and intermediates from within the EU are shown to be capable of preventing a *Renewable Pull*. The cost-reducing effect of imports on the energy system identified by Neumann et al. [25] can be confirmed and, beyond that, revealed as a decisive lever for the future competitiveness of Germany's chemical and steel industries.

For Japan and South Korea, follow-up import scenarios based on this study would be of interest to assess the extent to which targeted imports could also prevent the *Renewable Pull*. For both countries, the modeling of imports would require a detailed analysis of options beyond maritime transport, given their isolated geographic location.

## CONCLUSION

The *Renewable Pull* is a global phenomenon that affects not just a few nations but nine out of the top ten manufacturing countries of energy-intensive goods—Germany, the United States, Brazil, China, India, Japan, South Korea, Russia, and Iran—with Saudi Arabia being the only exception not affected.

Applying uniform capital costs, which primarily reflect energy system-related differences, Germany, Japan, South Korea, and India face the strongest *Renewable Pull*, while Saudi Arabia faces the lowest. Yet the inclusion of national capital costs has the potential to outweigh the differences in energy system costs. Under national capital cost assumptions, Saudi Arabia is no longer exposed to any *Renewable Pull* and Germany ranks second lowest, closely followed by the United States, while Iran, Russia, and India face the most intense *Renewable Pull*. The resulting energy system cost compositions of the energy system model—built on the complex interplay of hourly resolved national renewable potentials, salt cavern storage potentials, and electricity demand time series—demonstrate that there is no universal formula for a lowest-cost energy system composition for energy-intensive industries. It follows that differentiated calculations, such as those made in this study, are necessary to investigate the phenomenon of *Renewable Pull*.

Accounting for national capital costs, rather than relying solely on uniform or exemplary capital costs, has proven indispensable for meaningful results. However, unlike energy system modeling, the derivation of national capital costs is not physically grounded and remains influenced by political intervention, shifting financial assessments, natural hazards, or security risks. It turns out as a key strength of this study that—by showing all results calculated based on uniform capital costs and based on national capital costs—these two very differently rooted influences on the competitiveness of the energy-intensive industry can be observed separated from each other and, by that, are set into context with each other on a global scale.

The analysis across sectors further reveals that the *Renewable Pull* predominantly affects highly energy-intensive industries with low transport cost shares, such as chemical and steel, while cement remains largely unaffected. Accordingly, measures to mitigate the *Renewable Pull* should be sector-specific rather than spread across all industries.

Hydrogen and intermediate imports in the form of methanol and hot briquette iron emerge as highly effective mitigation measures, as demonstrated by the German import case examined in this study, for two reasons: first, hydrogen, hot briquette iron, and methanol can be imported at a lower cost than produced domestically in a country subjected to a *Renewable Pull*. In the case of Germany, imports from EU partners via truck and pipeline proved to be the most cost-effective options. Second, imports relieve the domestic renewable energy systems by reducing the overall electricity demand, thereby lowering the domestic renewable energy system costs, a finding confirmed by a sensitivity analysis in this study. Combined, these effects allow targeted imports to almost eliminate the *Renewable Pull* for Germany, even in a scenario where its currently low capital costs converge to higher international levels.

A clear recommendation follows for countries with good but limited renewable potential, such as Germany: ensuring the infrastructure for sufficient hydrogen and intermediate imports is essential to sustain industrial competitiveness in a renewable energy-powered future. Beyond their cost-reducing effect on domestic production, imports play a strategic role in guiding the expansion of renewable energy by ensuring that available low-cost potentials are not exceeded.

## METHODS

### *Industrial processes and the top manufacturing countries*

The relevance of energy costs varies significantly across subsectors within the manufacturing sector, which is internationally classified by the Standard Industrial Classification of All Economic Activities (ISIC) [32]. A useful indicator for assessing this relevance is the *Per Value Added Energy Intensity*, which expresses the relationship between *Value Added* and the energy required to achieve it. The energy intensity is notably high for three *ISIC* classes in particular, ranging between 5 and 25 megajoules per USD of value added (PPP 2015): *manufacturing of basic metals*, *manufacturing of other non-metallic mineral products*, and *manufacturing of chemical and pharmaceutical products* [2]. In other words, these sectors are among the most likely to be affected by a *Renewable Pull*.

Within the mentioned *ISIC* classes, representative products were selected, for which specific processes were modeled in this study. Given the wide range of products in the chemical industry, two exemplary processes were selected based on their particular relevance: the production of ammonia via the Haber–Bosch process and the production of olefins—a group of high-value chemicals including, for example, ethylene. High-value chemicals are used in the production of plastics, synthetic rubber, and fibers, whereas ammonia is primarily used for the production of synthetic fertilizers and pharmaceuticals and could also serve as an energy carrier in the future [33,34]. Both ammonia and olefins are considered primary chemicals, a category comprising high-value chemicals, ammonia, and methanol, which together account for approximately 67% of the total energy demand of the chemical sector. Conventionally, both processes are largely based on fossil natural gas [34]. However, future greenhouse gas-neutral alternatives exist for both, relying on green hydrogen as a feedstock [35]. Olefins consist of carbon and hydrogen. For the purposes of this study, the carbon source is assumed to be carbon dioxide that has been captured from the air using direct air capture technology. This enables a closed carbon cycle that is independent of the long-term disposition of the final product. The methanol-to-olefin route is applied, in which methanol is produced in a first step and converted into olefins in a second step. Methanol can also be provided as an intermediate [35,36]. Ammonia consists of nitrogen and hydrogen, with nitrogen supplied via an air separation unit. In both processes, the use of electrolysis, electrically supplied process heat, and power-based provision of other reaction inputs shifts the energy demand from fossil fuels to electricity. As an energetic substitute, the specific electricity demand exceeds that of conventional fossil-based production by more than a factor of fifty, depending on the process [35,37].

Steel is by far the most widely produced metal in the world. The highest demand comes from the construction, transportation, and mechanical engineering sectors [38,39]. Accordingly, this study investigates the process of primary steel production from the *ISIC* class *Manufacturing of Basic Metals*. In this process, iron ore is first reduced to iron via direct reduction and subsequently alloyed into steel in an electric arc furnace. *Direct reduced iron* can also be provided in the form of *hot briquetted iron* as an intermediate [40,41]. Hydrogen-based direct reduction is already being piloted today and is technologically similar to the established natural gas-based direct reduction process [42].

By volume, cement is the most widely used manufactured substance on Earth [43]. It therefore serves in this study as the representative process for the *ISIC* class *Manufacturing of Other Non-Metallic Mineral Products*. Due to process-related emissions, full defossilization of cement production is not possible in the medium term, despite initial approaches to alternative clinker materials [44]. Consequently, oxyfuel combustion combined with carbon capture and storage is selected as the decarbonization pathway for the cement process. Costs for $CO_2$ transport and storage are included. Within the model, $CO_2$ that cannot be captured via CCS must be removed from the atmosphere using direct air capture and subsequently stored to achieve a greenhouse gas-neutral balance.

The specific techno-economic assumptions used for each process, along with the underlying data sources, are documented in detail in Supplementary Excel File 1.

After defining the industrial products and processes to be examined in this study, the next step was to identify the countries with the highest production volumes of these products. Country rankings for each product were based on national production data for the steel and cement industries, and on sales data for the chemical industry [38,45]. Countries ranked from 1st to 10th received a descending score from 10 to 1 point per product. By summing these scores across all products, a list of the top 10 current *top* *manufacturing countries* was derived, including China, India, the United States of America, Russia, Saudi Arabia, Brazil, Germany, Japan, South Korea, and Iran.

### *The Model*

The results of this study are based on three models in total. All three models are based on a shared data foundation that includes techno-economic assumptions about technologies, renewable energy potentials, salt cavern storage potentials, and domestic electricity demands. Model 1 optimizes production costs for the 150 countries that have port access by treating each country as a single-node system. Shipping costs are added to the production costs for each country pair consisting of an export country and a *top manufacturing country*. The ratio of domestic production costs in *top manufacturing countries* to the import costs of finished products yields a *Renewable Pull Factor* for each country pair.

Based on Models 2 and 3, the extent to which import strategies can mitigate the *Renewable Pull* on Germany is investigated. Special attention is given to imports within the European Union (EU) to quantify the EU's potential as a single market to reduce the *Renewable Pull* on Germany. Model 2 optimizes the import of intermediates (HBI and methanol) via truck and hydrogen via pipeline resulting in corresponding import costs. It includes one node per EU member state and optimizes the annual import to Germany of 200 TWh of hydrogen, 10 Mt of methanol, or 10 Mt of HBI from within the EU.

| Import Options | Import Model | No | EU | Global |
|---|---|---|---|---|
| EU HBI Truck | 10 Mt Import from EU | x | o | o |
| EU Methanol Truck | 10 Mt Import from EU | x | o | o |
| EU Hydrogen Pipeline | 200 TWh Import from EU | x | o | o |
| EU HBI Ship | 1 Mt Import from EU | x | o | o |
| EU Methanol Ship | 1 Mt Import from EU | x | o | o |
| EU $LH_2$-Ship | 20 TWh Import from EU | x | o | o |
| Global HBI Ship | 10 lowest cost countries can provide 1 Mt each | x | x | o |
| Global Methanol Ship | 10 lowest cost countries can provide 1 Mt each | x | x | o |
| Global $LH_2$-Ship | 10 lowest cost countries can provide 20 TWh each | x | x | o |

EU Import Costs Model - Hydrogen:

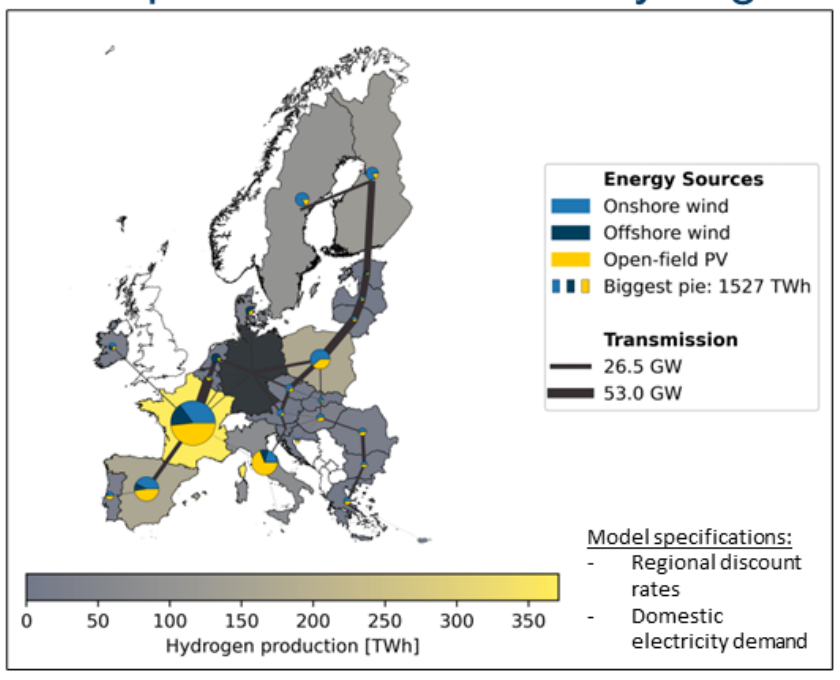

Shipping Tool:

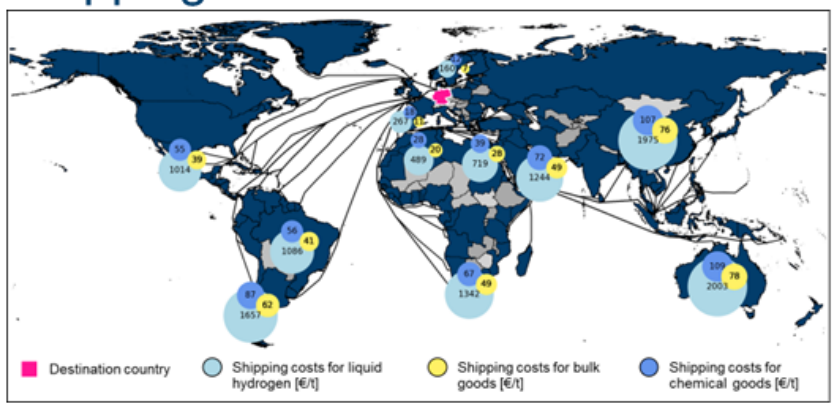

***Figure 10: Overview of import options for Model 1, o/x - import option possible or not***

Model 3 is a single-node model of Germany, analogous to the structure used in Model 1 for calculating the *Renewable Pull*. The key difference is that Model 3 includes the option to import the intermediates methanol and HBI, or hydrogen. These imports can originate either from within the EU – with costs derived from Model 2 – or globally via shipping from one of the 150 modeled countries, based on cost outputs from Model 1. To avoid dependency on a single country for global imports, import costs that can be undercut by at least ten countries according to Model 1 are considered.

Figure 11 presents the input data used for the energy system models developed within the ETHOS.fine framework [46,47]. The potentials for the electricity generation technologies are based on a comprehensive land eligibility analysis, combined with spatially disaggregated data on wind power and solar radiation [26], and on analyses conducted in collaboration with the *International Energy Agency* for the *Global Hydrogen Review 2024* [48]. With over 30 exclusion criteria and a spatial resolution of 100 meters, the dataset provides a high-quality basis regarding the availability of wind and solar energy for this study. Within the model it is applied as a set of time series representing theoretical availability (capacity factors between 0 and 1) for each generation technology, which the model can utilize by investing in limited capacity. The renewable potentials are clustered according to the full load hours: 10 clusters for wind technologies and 3 for open-field solar PV. This results in 23 electricity supply time series per country, totaling 3 450 electricity supply time series available to the optimization model.

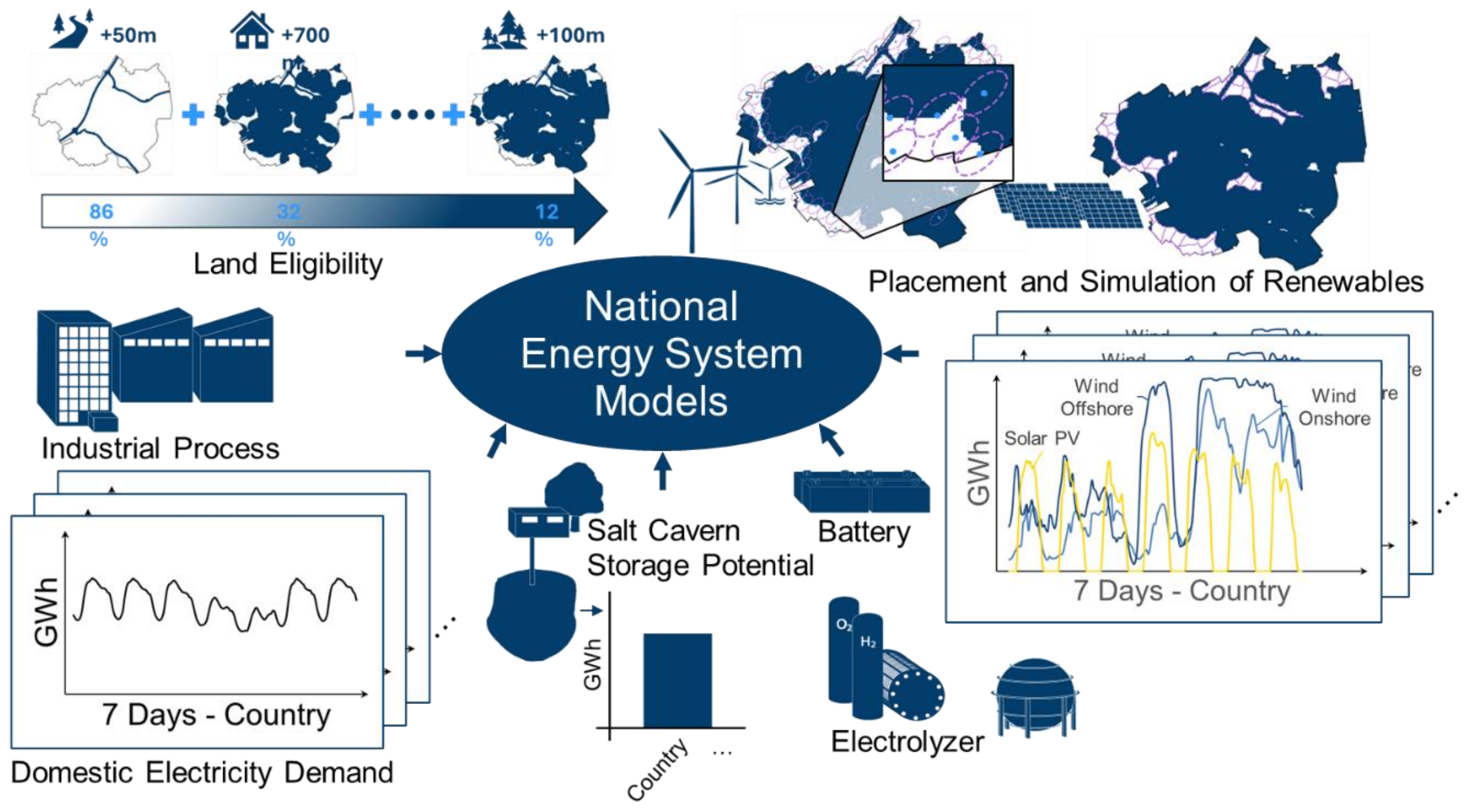

*Figure 11: The energy system model inputs*

The model includes large-scale battery storage, hydrogen tanks, and hydrogen salt cavern storage for countries with identified cavern storage potential. The national salt cavern storage potentials are derived from a detailed global geological suitability and land eligibility assessment, as presented by Franzmann et al., currently available as a preprint [49]. National electricity demand projections for the year 2050 are based on results from the Integrated Assessment Model GCAM 5.3, using a Net Zero 2050 scenario [19,50]. This scenario builds on the assumptions of the *middle of the road scenario* of the Shared Socioeconomic Pathway II, including projections for population growth and GDP development, under the premise that the 2°C climate target can be met [18]. The national hourly electricity time series are generated by combining the scenario-specific annual electricity demand projections with hourly load curves from PLEXOS [51]. For the single-node Models 1 and 3, the resulting electricity demand was reduced by half. This adjustment was made for two reasons: first, not all renewable technologies contributing to a complete power mix—such as hydropower, geothermal energy, rooftop PV, or biomass—are included on the supply side; second, to avoid double-counting the industrial sector. The impact of an increasing electricity demand is shown within the supplementary information Figure S4 15. The domestic electricity demand and the industrial production are jointly solved within one optimization problem. In the post-processing step, total system costs are allocated proportionally to the electricity consumption for domestic demand and for the industrial production, including the hydrogen and intermediate production. This enables the calculation of cost per ton of the respective industrial product within a fully integrated national energy system, in which neither domestic electricity demand nor industrial production is granted preferential access to the most favorable renewable potentials. This stands in contrast to other studies presented in the introduction, where local demand is subtracted from available potentials beforehand—if considered at all—thereby forcing the modeler to intervene and implicitly prioritize one sector over the other. In the integrated approach applied in this study, both sectors are optimized jointly within the same system framework, which could be interpreted as an approximation of a macroeconomic view. To reduce computational time, all time series were aggregated prior to optimization using the *Time Series Aggregation Module* (TSAM), resulting in 40 typical days with 24 time steps each [47].

The specific techno-economic assumptions used for each process within the modeled energy system, along with the underlying data sources, are documented in detail in Supplementary Excel File S1.

### *Shipping costs*

In order to transport products, intermediates, and liquid hydrogen by ship, numerous decisions must be made regarding transportation routes and costs. The distances from port to port for each country combination are calculated using the searoute-py python package [52]. The port selection is based on the ability to handle large ship dimensions. The primary criterion is the maximum allowable length of the ship. If that is unavailable, the maximum breadth or draught is used instead. For countries with multiple ports, one port per cardinal direction is selected based on these parameters. In some cases, manual corrections are made due to geographic factors such as canal tolls or excessive detours [53]. To enable a comprehensive transportation cost assessment, the calculation incorporates multiple cost parameters. The ship type is selected based on the transported cargo. Average values for investment cost, lifetime, and cargo capacity are derived from a global ship database [54]. Additional parameters, including maintenance [55], insurance [56], labor [56], canal tolls [57,58], port days [59,60], and port fees [61] are sourced from the literature. The specific techno-economic assumptions used for the ship and the liquefaction unit, along with the underlying data sources, are documented in detail in Supplementary Excel File S1.

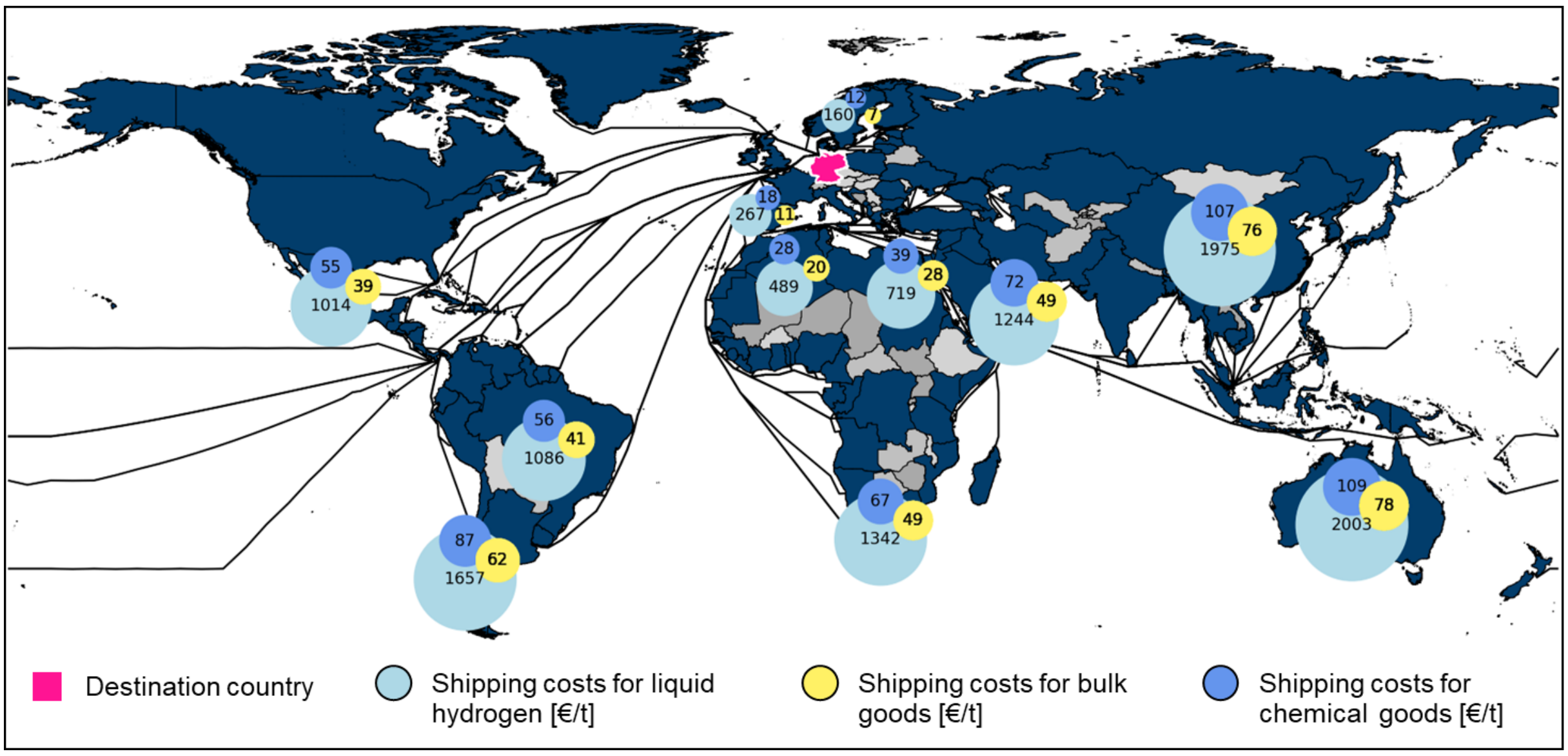

*Figure 12: Exemplary shipping costs to Germany*

### *Raw material costs*

This study considers the raw materials limestone, iron ore, and scrap. Additionally, the fossil energy sources coal, natural gas, and oil are included for cement production. To the authors' knowledge, no freely available global database for national commodity prices currently exists. To address this gap, a global database of nationally differentiated raw material costs based on the public available *UN Comtrade Database* was created [62,63]. A decision logic was set, where the central element is the weighing of export and import prices according to export and import mass flows for each country-material combination [63]. Therefore, averaged entries over the period 2015-2019, from the UN Comtrade database are used [62]. The post-processing includes the correction of outliers that could potentially arise due to errors in reporting to the UN Comtrade database [64]. A *median absolute deviation* of 2 is used to set a reasonable upper and lower limit. For the country-material combinations of this study, a total of 31.2% of the data points had to be corrected by the defined limits. The entire database generated can be found in Supplementary Excel File S1.

### *Country-specific capital costs*

Capital costs are represented in the energy system modeling through the discount rate. Using publicly available financial market data and a transparent methodology, Damodaran et al. determined national discount rates for 156 countries [65]. Egli et al. have already applied these rates in the context of assessing the competitiveness of African hydrogen imports to Europe [66]. However, the increasing risk of natural hazards in the future has not yet been explicitly accounted for [67]. Since this study uses 2050 as its base year, the capital costs were derived using the approach outlined by Stargardt et al. in their preprint, which incorporates a natural hazard-adjusted discount rate [28]. This rate incorporates Damodaran et al.'s financial market-based discount rates and a natural hazard risk component at a ratio of 75:25, respectively. Compared to purely financial, market-based interest rates, this adjustment benefits countries in Africa, the Middle East, and the region between Europe and Pakistan, as well as eastern Latin America. In contrast, discount rates increase for North America and most countries in Central, South, and Southeast Asia [28].

### *Defining the Renewable Pull*

In the existing literature, the concept of the *Renewable Pull* has been used to describe a theoretical future phenomenon: the deindustrialization of certain regions due to superior renewable energy potentials and consequently lower energy costs in other regions [7].

Equation 1

$$Renewable\ Pull = \frac{[\frac{€}{t}]_{domestic\ production_i}}{[\frac{€}{t}]_{import_e}} - 1$$

$$i = importing\ country, e = exporting\ country$$

$$\text{REP} = Renewable\ Pull$$

$$\text{REP_Max} = \text{Maximal } Renewable\ Pull \text{ of 1 region}$$

$$\text{REP_20} = Renewable\ Pull \text{ exceeded by 20 regions}$$

This study takes the concept one step further by quantifying the *Renewable Pull* between two countries using a defined Equation 1. A resulting value greater than zero indicates that a *Renewable Pull* exists between the respective country combination. For each of the 10 *top manufacturing countries* considered in this study, the *Renewable Pull* exerted by each of the 150 modeled countries is determined. Based on that, two additional indicators are derived from this analysis: the *REP_20* value, which reflects the strength of *Renewable Pull* exceeded by the 20 lowest cost export countries, and the *REP_MAX* value, which represents the maximum *Renewable Pull* exerted by a single export country. The renewable pull, potentially leading to the so-called green-relocation is an extreme scenario. In this study the renewable pull was defined as a number and can also be seen as a measure of future competitiveness of an industrial location in global trade.

One note should be made regarding the interpretation of this study's results. The *Renewable Pull* in this work is assessed based on energy system costs. In real industrial processes, however, energy costs are derived from electricity prices. If Germany can be cost-competitive from an energy system cost perspective, the question remains whether electricity prices will also permit competitiveness. Currently, up to 55% of the electricity price for industrial end users in Germany consists of grid charges, taxes, levies, and duties [13,29,68,69]. In order to be competitive, the levy system must therefore allow for it. From an energy system perspective, including targeted imports, Germany can provide a competitive renewable electricity supply for the industry.

## RESOURCE AVAILABILITY

### *Lead contact*

Requests for further information should be directed to and will be answered by the lead contact, Arne Burdack (a.burdack@fz-juelich.de)

### *Materials availability*

This study did not involve any physical materials.

### Data and code availability

A dataset of the model results, including results that go beyond the results shown, will be made available on Zenodo after peer review. All techno-economic assumptions can be found in the supplementary information. Input data used based on complex preprocessing workflows are described in detail within the methodology. The energy system model framework ETHOS.FINE is open source and available under https://github.com/FZJ-IEK3-VSA/FINE [46].

## ACKNOWLEDGEMENTS

This work was partly funded by the European Union (ERC, MATERIALIZE, 101076649). Views and opinions expressed are, however, those of the authors only and do not necessarily reflect those of European Union or the European Research Council Executive Agency. Neither the European Union nor the granting authority can be held responsible for them.

This work received partial funding through the project *Follow-ETSAP* (grant number 03EI1064), supported by the Federal Ministry for Economic Affairs and Climate Action on the basis of a decision by the German Bundestag.

This work was partly supported by the Helmholtz Association under the program “Energy System Design.”

## AUTHOR CONTRIBUTIONS

A.B.: Conceptualization, Formal analysis, Investigation, Methodology, Software, Validation, Visualization, Writing – original draft, Writing – review & editing

M.S.: Investigation, Software, Visualization, Methodology – Shipping costs

C.W.: Software, Visualization

K.K.: Investigation

D.S.: Supervision

J.L.: Supervision, Writing – review & editing

H.H.: Supervision, Writing – review & editing, Conceptualization

## DECLARATION OF INTERESTS

The authors declare no competing interests.

## DECLARATION OF GENERATIVE AI AND AI-ASSISTED TECHNOLOGIES IN THE WRITING PROCESS

During the preparation of this work, the authors used *DeepL Write*, *ChatGPT*, and *LanguageTool* to improve the readability and language of the manuscript. After using these services, the authors reviewed and edited the content as needed and take full responsibility for the content of the published article.

## SUPPLEMENTAL INFORMATION

Excel file S1 – Underlying techno-economic assumptions

Supplementary figures and tables

*Table S1 1: Model workflow of the three models of the study*

| Model | Objective | Data | Optimization | Postprocessing | Postprocessing |
|---|---|---|---|---|---|
| | | | Production | Transport | Import |
| **1** | 1) Domestic & export country production<br>2) *Renewable Pull*<br>3) Intermediate import | Techno-economic parameters: specific energy demands, cost assumptions, efficiencies<br><br>Renewable potentials: solar PV, wind onshore, wind offshore<br><br>Large-scale storage potentials: salt cavern<br><br>Demand time series: domestic electricity demand | **Commodity production** costs with a one-node energy system model based on the ETHOS.FINE framework | **Shipping** costs of commodity with Shipping-Tool | ***Renewable Pull Factor*** as a ratio between product import and domestic production |
| Model workflow<br>►►►►►►►► | | | Optimization | | Postprocessing |
| | | | Production | Transport | Import |
| **2** | German intermediate & hydrogen import from EU | | **Commodity production** with a multi-node energy system model based on the ETHOS.FINE framework | - **Pipeline** hydrogen transmission<br>- Methanol, or hot briquetted iron **truck** transport | **Import costs** of hydrogen, methanol, or hot briquetted iron from the EU |
| Model workflow<br>►►►►►►►► | | | Optimization | | Postprocessing |
| | | | Import | Production | Costs |
| **3** | German production costs with imports | | Methanol, hydrogen, hot briquetted iron<br>By ship: **Globally**<br>By Truck and Pipeline: **From the EU** | **Commodity production** costs with a one-node energy system model based on the ETHOS.FINE framework | **Domestic production costs** |

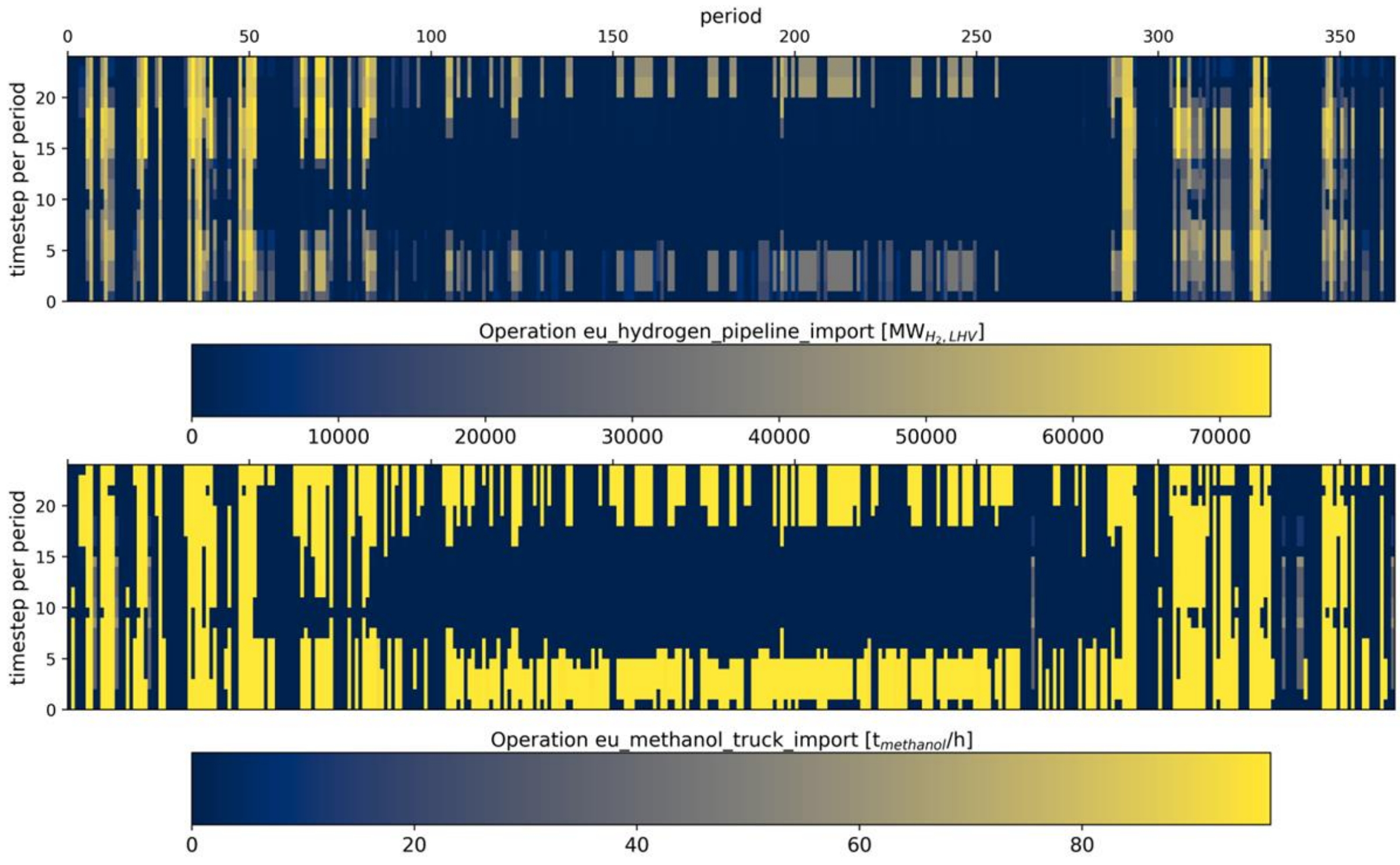

*Figure S2 13: Optimized import operation, model specifications: National Capital Costs*

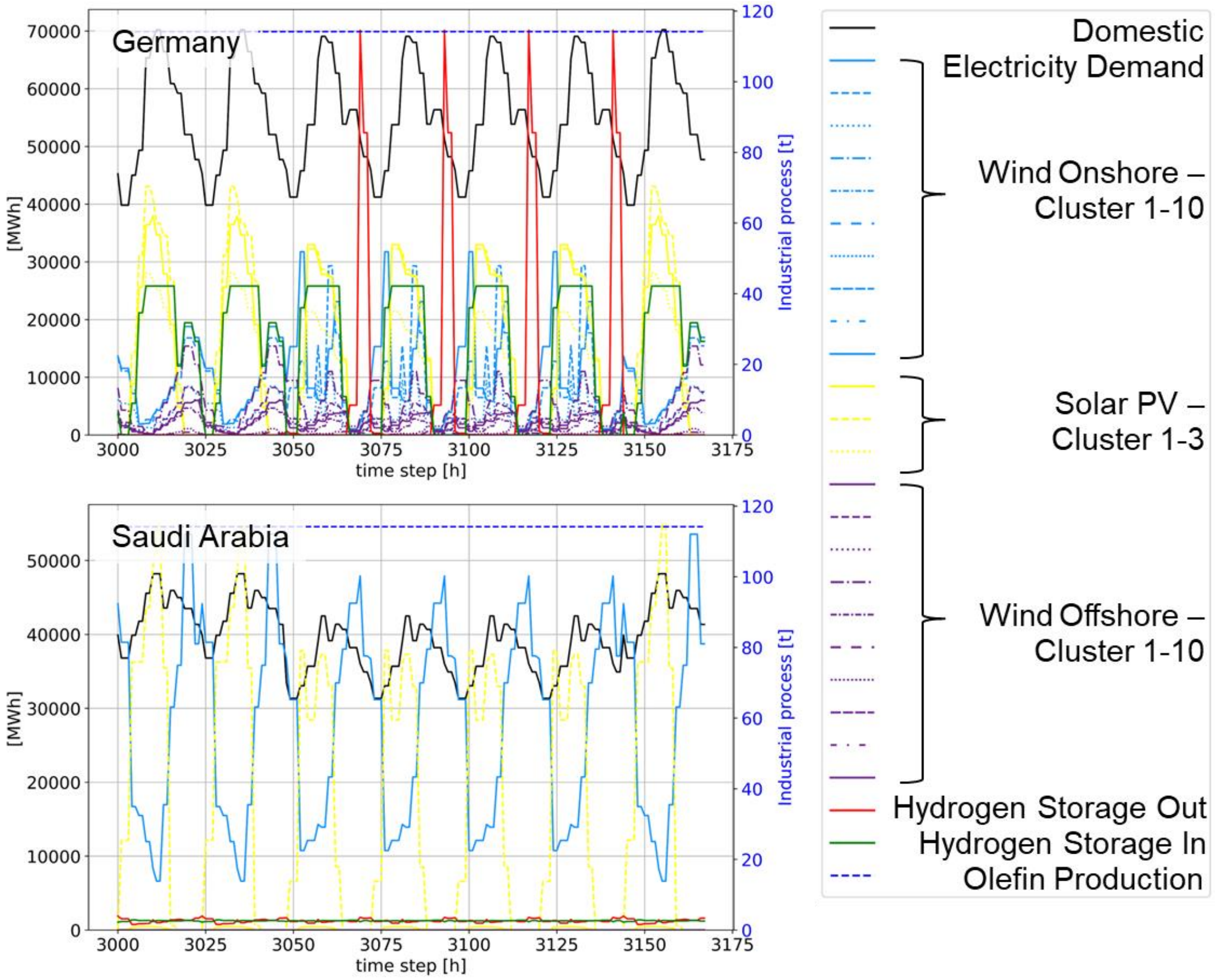

*Figure S3 14: Optimized operation of electricity supply, hydrogen storage, and final demands over a one-week snapshot within a year*

Figure S3 14 shows two countries with very different renewable potentials and domestic electricity demands.

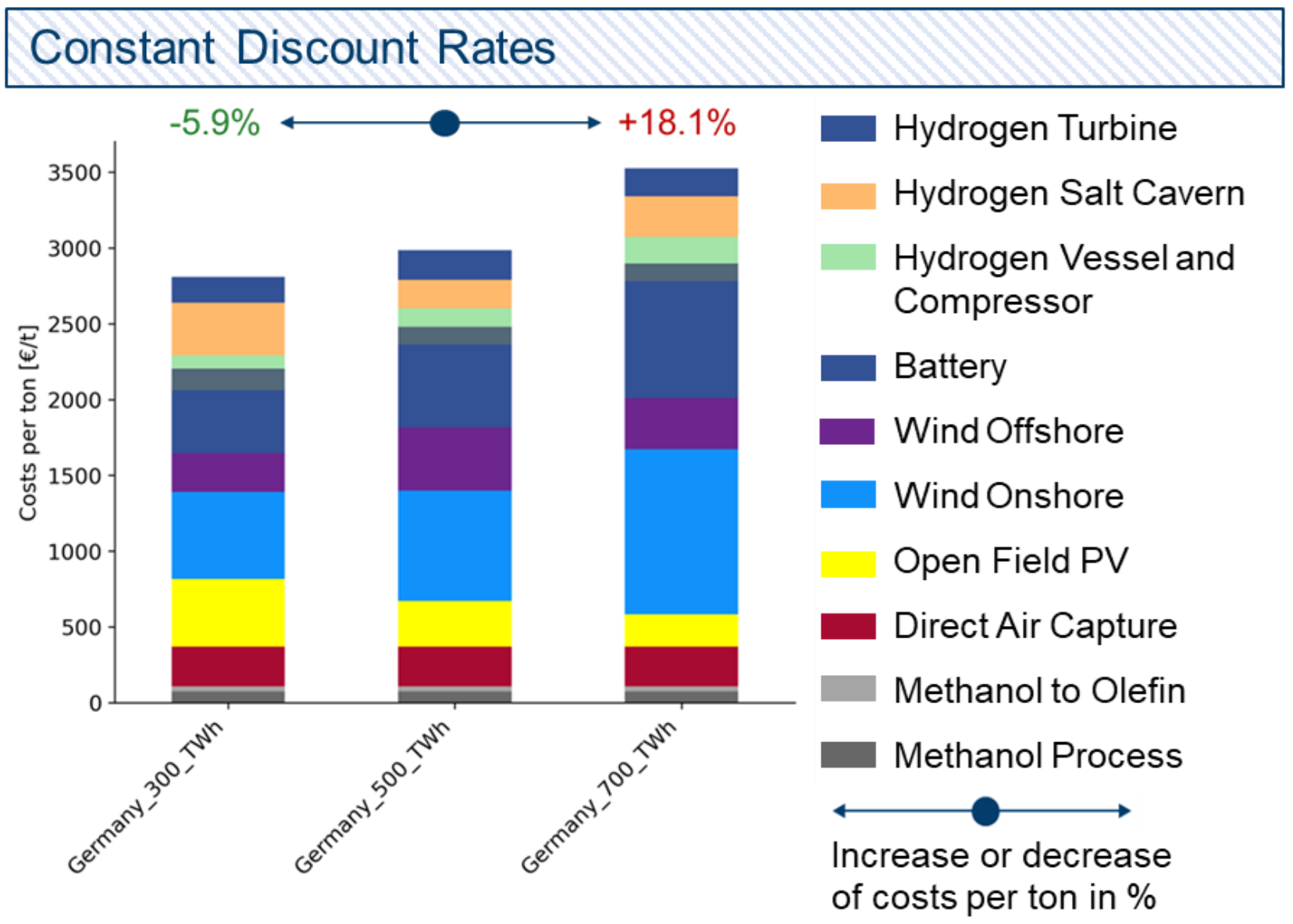

*Figure S4 15: Sensitivity of olefin production costs in the optimized German energy system to electricity demand*